\documentclass[prd,aps,twocolumn,a4paper,floatfix]{revtex4}

\usepackage{graphicx,psfrag}
\usepackage{mathrsfs}
\usepackage{amsmath,amsfonts,amssymb}
\usepackage{multirow}
\usepackage{comment}
\usepackage{hyperref}
\usepackage{bbm}
\usepackage[utf8]{inputenc}
\usepackage{placeins}
\usepackage[normalem]{ulem}
\usepackage{slashed} 
\usepackage{amsthm}
\usepackage{latexsym}
\usepackage[dvips]{epsfig}
\usepackage{wasysym}
\usepackage{bm}
\usepackage{yhmath} 
\usepackage{wrapfig}
\usepackage{stmaryrd}
\usepackage{pifont} 
\usepackage{slashed}

\def\p{\partial}
\def\snabla{\slashed{\nabla}}
\def\sDelta{\slashed{\Delta}}

\allowdisplaybreaks

\begin{document}

\title{A semi-linear wave model for critical collapse}

\author{Isabel Su\'arez Fern\'andez, Rodrigo Vicente and David Hilditch}

\affiliation{
CENTRA, Departamento de F\'isica, Instituto Superior
T\'ecnico IST, Universidade de Lisboa UL, Avenida Rovisco Pais 1, 1049
Lisboa, Portugal}

\date{\today}

\begin{abstract}
In spherical symmetry compelling numerical evidence suggests that in
general relativity solutions near the threshold of black hole
formation exhibit critical behavior. One aspect of this is that
threshold solutions themselves are self-similar and are, in a certain
sense, unique. To an extent yet to be fully understood, the same
phenomena persist beyond spherical symmetry. It is therefore desirable
to construct models that exhibit such symmetry at the threshold of
blow-up. Starting with deformations of the wave equation, we discuss
models which have discretely self-similar threshold solutions. We
study threshold solutions in the past light cone of the blow-up
point. In spherical symmetry there is a sense in which a unique
critical solution exists. Spherical numerical evolutions are also
presented for more general models, and exhibit similar behavior. Away
from spherical symmetry threshold solutions attain more
freedom. Different topologies of blow-up are possible, and even
locally the critical solution needs reinterpretation as a
parameterized family.
\end{abstract}

\maketitle

\section{Introduction}\label{Section:Introduction}

The veracity of the weak and strong cosmic censorship conjectures is
of monumental importance in classical~$3+1$ dimensional gravity. Of
these, the weak cosmic censorship conjecture can be thought of
informally as the statement that, given generic asymptotically flat
initial data, the resulting solution will exist globally outside of a
black hole region~\cite{Pen69,Chr99}. Strong cosmic censorship is
likewise the conjecture of uniqueness of solutions emanating from
generic initial data, and is directly related to regularity of
solutions at blow-up. A natural role for numerical relativity in this
context is in the construction of potential counterexamples. A hope
might be to give convincing evidence that an open set of initial data
do not have complete black hole exteriors, or are sufficiently regular
at blow-up so that they may be extended non-uniquely.

\begin{figure}[t!]
\centering
\includegraphics[width=0.47\textwidth]{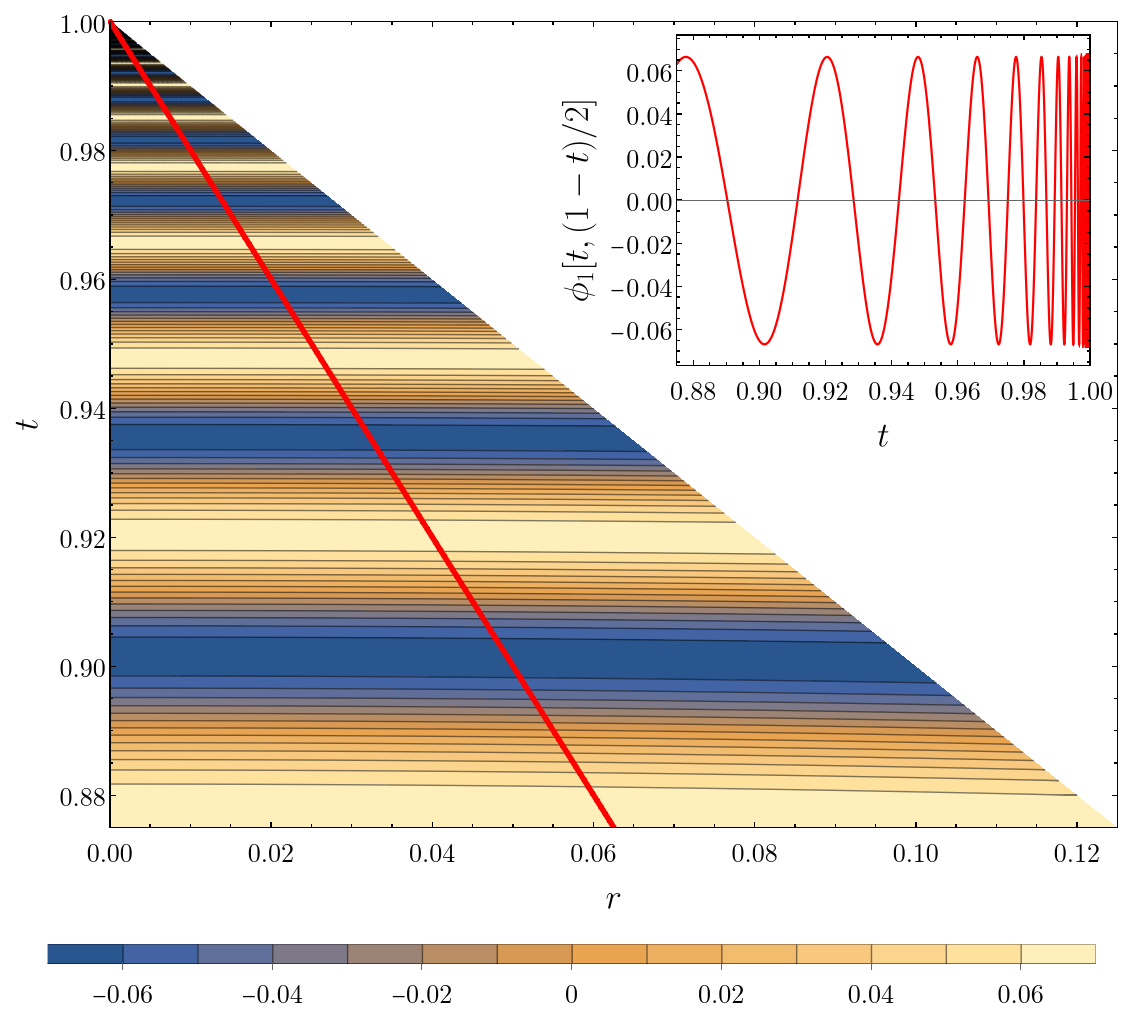}
\caption{A contour plot of a threshold solution of model~3
  with~$A_3=1/15$, see Eqn.~\eqref{model3_D}. The parameter here was
  chosen simply for clarity of plotting. This threshold solution blows
  up in~$H^1$ (and not in~$L^2$) in a discretely self-similar fashion
  at~$(t_\star,r_\star)=(1,0)$. In the inset the solution is plotted
  along the red curve~$t+2r(t)=1$ indicated in the main plot. The
  figure is naturally compared with Fig.~$1$ of~\cite{ReiTru19}.}
\label{Fig:Critical}
\end{figure}

One strategy to try and construct such {\it extreme} spacetimes is the
following: consider a one-parameter family of initial data such that
small values of the (strength) parameter result in data close to
flat-space, with larger values being more and more deformed. Then tune
that strength parameter to the threshold of black hole
formation. Starting with the pioneering work of Choptuik~\cite{Cho93},
studies along these lines in spherical symmetry revealed behavior
which has since come to be known as critical phenomena in
gravitational collapse~\cite{GunGar07}. In short, it has been found
that for a given family there is, in a sense, a single solution lying
between dispersion and collapse to form a black hole. Numerical
evidence suggests that these solutions have naked singularities, but
these are not considered true counterexamples to the cosmic censorship
because the phenomena occurs only by fine-tuning the initial
data. These threshold solutions are fascinating. Empirically, they are
either continuously or discretely self-similar, and, remarkably, for a
given model appear to be unique, in the sense that all families of
initial data exhibit the {\it same} threshold solution, called the
critical solution, and are thus in some sense as close as possible to
being attractors in solution space. Consequently, when considered as a
function of the strength parameter, solutions naturally give rise to
power law behavior near the threshold. For example, when the critical
solution is discretely self-similar, the maximum of any non-vanishing
curvature scalar, viewed as a function of distance from the threshold
in phase space, follows a power law with a superposed periodic
wiggle~\cite{Gun96a,HodPir97,GarDun98}. Beyond spherical symmetry
similar behavior has also been observed, although typically with
features that have yet to be persuasively explained. For example in
the collapse of electromagnetic waves~\cite{BauGunHil19}, threshold
solutions appear to be only {\it approximately} self-similar.

Much of the picture of critical collapse described above was formed
through a combination of the empirical findings of numerical studies
and thoughtful heuristic modeling. To understand what might be shown
rigorously it is therefore desirable to construct {\it maximally}
simple models that capture the qualitative behavior near the threshold
of blow-up. Various such investigations have been made in the
literature~\cite{Lie02,BizChmTab04,Lie05a,BizMaiWas07,BizZen08}, but
all so far exhibit continuous, rather than discrete
self-similarity. From the point of view of nonlinear PDEs we therefore
seek a simple system that admits a small-data global existence result,
but with large data breaking down, and a unique {\it discretely}
self-similar critical solution at the threshold between the two
regimes. Illustrative would furthermore be if, just as the
system,
\begin{align}
\square\phi=\nabla_a\phi\nabla^a\phi\,,
\end{align}
can be used to motivate the utility of the classical null
condition~\cite{Kla80}, the model were to indicate the structural form
of nonlinear terms that generate self-similar critical solutions. We
denote by~$\nabla$ the Levi-Civita derivative compatible
with~$\eta_{ab}$, the Minkowski metric, and~$\square$ the flat space
d'Alembertian. The first aim of this work is to give, for the first
time, such a model. In what follows we therefore present a number of
different toys. With the simplest parameter choice, one of our models
is
\begin{align}
  \square\phi+\frac{\phi-\sqrt{1-\phi^2}}{1-\phi^2}
  \nabla_a\phi\nabla^a\phi=0\,.\label{eqn:dss_model_simple}
\end{align}
In Figure~\ref{Fig:Critical} we plot a spherical solution to the
associated model equation at the threshold of blow-up (albeit with
slightly different parameters for the purposes of plotting) in the
past light-cone of the blow-up point.

We are furthermore interested in the properties of solutions near
blow-up and in the status of conjectures related to cosmic censorship
both in and beyond spherical symmetry. All of the models we study,
like~\eqref{eqn:dss_model_simple}, are semi-linear or equivalent to a
semi-linear PDE. That is, there is no nonlinearity whatsoever in the
principal part. The principal part is furthermore taken to be the
d'Alembert operator associated with the Minkowski metric. With the
metric so fixed, there are no notions of either trapped surface, or
black hole formation intrinsic to the model. Therefore, in any setup
with this restriction, if blow-up is present without fine-tuning
initial data, the obvious conjecture directly analogous to weak cosmic
censorship must be false. Our focus is instead on the nature of
solutions at the threshold of blow-up, and the extent to which
critical behavior is obtained for solutions nearby in phase space.

The paper is structured as follows. In
section~\ref{Section:Self-similar} we discuss different notions of
blow-up and self-similarity. Then in section~\ref{Section:Models} we
explain the construction of our various models. In
section~\ref{Section:Critical} we study the behavior of threshold and
near-threshold solutions. Afterwards, we restrict to spherical
symmetry and present a set of numerical evolutions in
section~\ref{Section:Numerics}. We conclude in
section~\ref{Section:Conclusion}.

\section{Self-similar functions}\label{Section:Self-similar}

Solutions to our models either live forever or terminate at some
finite time. The manner in which solutions blow up in our models splits
into two categories depending on the model. Either just first
derivatives, or the field itself explodes pointwise. Each of these has
an obvious, although inequivalent, analogue in~$L^2$-like norms.
Solutions at the threshold of blow-up may exhibit more structure,
described in many cases by self-similarity, a special class of scale
invariance. Likewise, in our examples two types of self-similarity,
discrete and continuous, manifest. Since these threshold solutions are
just examples of blow-up, there must then be a relationship between
these notions, which we discuss in this section.

\paragraph*{\textbf{Notions of blow-up.}} The choice of a function space
in which mathematical results are formulated and proven is subtle, but
for our models a simple overview will suffice. A function~$f(t,x^i)$
is said to be in~$L^2$ at instant~$t$ if the integral
\begin{align}
  ||f||_{L^2}\equiv\left(\int_{\mathcal{D}[f(t,\cdot)]} \textrm{d}\Sigma
  \left|f(t,x^i)\right|^2\right)^{1/2}
\end{align}
exists and is finite, where~$\mathcal{D}[f(t,\cdot)]$ is the domain
of~$f(t,x^i)$. Throughout, the coordinates~$t,x^i$ are taken to be
global inertial on Minkowski and~$\textrm{d}\Sigma$ denotes the
natural volume form induced in level sets of~$t$. Another norm that
appears in the study of wave equations is given by
\begin{align}
  &||f||_{H^1}\equiv \nonumber \\
  &\left(\int_{\mathcal{D}[f(t,\cdot)]} \textrm{d}\Sigma\,
  \Big(\left|f(t,x^i)\right|^2 +\sum_i\left|\p_i f(t,x^i)\right|^2\Big)
  \right)^{1/2}\,.
\end{align}
When this quantity is finite we will colloquially refer to the
function as being~$H^1$. Here and in what follows~$\p_i$ denotes the
partial derivative~$\p/\p{x^i}$. More generally, a function~$f(t,x^i)$
is said to be in the Sobolev space~$H^k$ at instant~$t$ if the
norm~\cite{Rin09b}
\begin{align}
	||f||_{H^k}\equiv\left(\int_{\mathcal{D}[f(t,\cdot)]} \textrm{d}\Sigma
	\sum_{|\alpha|\leq k}\left|\p^\alpha_{i} f(t,x^i)\right|^2\right)^{1/2}
\end{align}
is finite. In the last expression, we are using the multi-index
notation with the $3$-tuple~$\alpha=(\alpha_1,\alpha_2,\alpha_3)$ of
non-negative integers, where~$|\alpha|=\alpha_1+\alpha_2+\alpha_3$
and~$\p_i^\alpha
f\equiv\p_x^{\alpha_1}\p_y^{\alpha_2}\p_z^{\alpha_3}f$. Note
that~$H^0\equiv L^2$. We additionally consider
the~$E^1$-norm~\cite{Kla80}
\begin{align}
&||f||_{E^1}\equiv \nonumber \\ &\left(\int_{\mathcal{D}[f(t,\cdot)]}
  \textrm{d}\Sigma\, \Big(\left|\p_t f(t,x^i)\right|^2 +\sum_i
  \left|\p_i f(t,x^i)\right|^2\Big)\right)^{1/2}\,,
\end{align}
which is perhaps the norm that appears most naturally for wave
equations. When a function which is initially in~$H^k$ ($E^1$) fails
to be so at some instant~$t'$, we say that it ``blows up'' in~$H^k$
($E^1$) at that instant~$t'$. Clearly, if a function blows up
in~$H^k$, it blows up also in~$H^{k'}$ with~$k'>k$.

We might like to restrict our attention exclusively to classical
solutions with bounded derivatives (called~$C_b^k$) and discuss
blow-up exclusively in terms of the field or derivatives
thereof. Alternatively, we may want to consider the function
space~$L^\infty$ of measurable bounded functions; this space contains
and has the same norm of~$C_b^0$.  However, besides the inconvenient
fact that proofs of existence and so forth do not naturally appear in
these spaces, the formulation of the weak cosmic censorship
itself~\cite{Chr99} is given in terms of local~$L^2$ integrability of
the connection coefficients. Intuitively this makes sense, because we
can introduce local inertial coordinates at any point, and so if some
blow-up occurs and is unavoidable we might expect it to be associated
with at least one derivative of the metric, and hence the connection
appears naturally. From a modeling point of view, we are therefore
more interested in finding semi-linear wave equations with blow-up
in~$E^1$ rather than in~$L^2$.

\paragraph*{\textbf{Self-similar functions.}} The notion of self-similarity
has to do with invariance under certain scale transformations. We
consider two kinds of self-similarity; continuous (CSS) and discrete
(DSS). A scalar function~$f$ is said to be CSS if there exists a
coordinate system~$(t,x^i)$ and a~$\nu \in \mathbb{R}$ such that
\begin{align} 
  f(\lambda t, \lambda x^i)=
  \lambda^\nu f(t, x^i)\,,
  \label{self-sim}
\end{align}
for any~$\lambda>0$. Notice that we have chosen our coordinates so
that the origin coincides with the center of the symmetry. When~$\nu$
is an integer, $f$ is also called a homogeneous function of
degree~$\nu$. On the other hand, a function~$f$ is said to be DSS if
there exists coordinate system~$(t,x^i)$, a~$\nu \in \mathbb{R}$, and
{\it some}~$\Delta>0$ such that~\eqref{self-sim} holds
for~$\lambda=e^{-m \Delta}$, with any~$m\in \mathbb{Z}$. Thus, DSS
functions have a fractal-type behavior under scale
transformations. The condition~\eqref{self-sim} is often expressed in
\textit{similarity coordinates}~$(T,X^i)=(-\log |t|, x^i/t)$ as
\begin{align}
	f(T+\tau,X^i)=e^{-\nu \tau} f(T,X^i)\,,
\end{align}
where $\tau=-\log \lambda$. In the case of DSS functions the last
condition is satisfied for~$\tau=m \Delta$.

\paragraph*{\textbf{Self-similarity and blow-up.}} Interestingly,
self-similar functions offer special examples of blow-up, either
pointwise or under some integral norm. For example, a CSS function
satisfying~\eqref{self-sim} -- with~$x^i$ the canonical Cartesian
coordinates -- satisfies also
\begin{align} \label{CSSHk}
&\int_{\mathbb{R}^3} \textrm{d}\Sigma
\sum_{|\alpha|=k}\left|\p^\alpha_i f(t,x^i)\right|^2
=\nonumber\\ &\frac{1}{\lambda^{2(\nu-k+3/2)} }
\int_{\mathbb{R}^3} \textrm{d}\Sigma \sum_{|\alpha|=k}
\left|\p^\alpha_i f(\lambda t,x^i)\right|^2,
\end{align}
with any~$\lambda$. Here we are assuming that the domain of~$f$
is~$\mathbb{R}^3$, except (possibly) a set of zero measure. Choosing
$\lambda=1/|t|$ we see that for~$t<0$
\begin{align}
	&\int_{\mathbb{R}^3} \textrm{d}\Sigma
	\sum_{|\alpha|=k}\left|\p^\alpha_i f(t,x^i)\right|^2
	=\nonumber\\
        &\frac{1}{|t|^{2(k-\nu-3/2)} }\int_{\mathbb{R}^3}
        \textrm{d}\Sigma \sum_{|\alpha|=k}
	\left|\p^\alpha_i f(-1,x^i)\right|^2,
\end{align}
Thus, a nontrivial CSS function with~$\nu \leq -3/2+k$ cannot be
in~$H^k$ for all times; if it is in~$H^k$ at a particular
instant~$t<0$, it must blow up at~$t=0$. It is easy to see that the
same argument goes through for CSS functions in~$E^1$; if a
non-trivial CSS function with~$\nu\leq -1/2$ is in~$E^1$ at a
particular instant~$t<0$, it must blow up at~$t=0$. A CSS function
satisfies
\begin{align} \label{CSSCk}
  \p_{t}^{\alpha} f(t,x^i)&=\frac{1}{\lambda^{\nu-k}}
  \,\p_{\lambda t}^{\alpha} f(\lambda t,\lambda x^i)\,, \qquad |\alpha|=k\,,
  \nonumber \\ 
  \p_{x^i}^{\alpha} f(t,x^i)&=\frac{1}{\lambda^{\nu-k}}
  \,\p_{\lambda x^i}^{\alpha} f(\lambda t,\lambda x^i)\,, \qquad |\alpha|=k\,.
\end{align}
with any~$\lambda$. Choosing again~$\lambda=1/|t|$, we obtain
\begin{align} 
\p_{t}^{\alpha} f(t,x^i)&=\frac{1}{|t|^{k-\nu}}
\,\p_{\lambda t}^{\alpha} f(\lambda t,\lambda x^i)_{|(-1, x^i/|t|)}\,,
\quad |\alpha|=k\,,
\nonumber \\ 
\p_{x^i}^{\alpha} f(t,x^i)&=\frac{1}{|t|^{k-\nu}}
\,\p_{\lambda x^i}^{\alpha} f(\lambda t,\lambda x^i)_{|(-1, x^i/|t|)}\,,
\quad |\alpha|=k\,.
\end{align}
So, a CSS function with~$\nu<k$ cannot be in~$C_b^k$ for all times; if
it is in~$C_b^k$ at a particular instant $t<0$, it must blow up
at~$t=0$.  The two arguments above can be easily extended to DSS
functions with the same bounds on~$\nu$ by taking the limit~$t \to
0^-$ through a sequence~$t_m=-1/\lambda_m=-e^{-m \Delta}$. Note that
DSS functions satisfy~\eqref{CSSHk} and~\eqref{CSSCk} for a discrete
set of values of~$\lambda$. The results of this section are summarized
in Tab.~\ref{Tab:BlowUp_Cond}.

\begin{table}[t]
	\centering
	\begin{tabular}[t]{ c | c | c }
		\hline\hline
		  $H^k$ & $E^1$ & $C_b^k$  \\ \hline
		 $\nu\leq -3/2+k$ & $\nu \leq -1/2$ & $\nu <k$  \\
		\hline\hline
	\end{tabular}
	\caption{A CSS or DSS function with degree~$\nu$ (see
          Eqn.~\eqref{self-sim}) must blow up in a given function norm
          (first line) if the associated condition in~$\nu$ (second
          line) is satisfied.
		\label{Tab:BlowUp_Cond}}
\end{table}

\paragraph*{\textbf{Sobolev embedding.}} It can be shown (for details
see Theorem~6.5 in~\cite{Rin09b}) that, for each~$k$ a non-negative
integer and~$s>k+3/2$, there is a constant~$c$ such that
\begin{align}
	||f||_{C_b^k}\leq c \,||f||_{H^s}\,,
\end{align}  
with~$f$ an arbitrary function. In
particular, for~$k=0$ and~$s=2$,
\begin{align} \label{SobolevEmbed}
	||f||_ {C_b^0}\leq c\, ||f||_{H^2}\,.
\end{align}
This implies that if a function blows up in~$C_b^0$, it also blows up
in~$H^2$. Since the~$C_b^0$-norm is equal to the~$L^\infty$-norm, if a
function blows up in~$L^\infty$ it also blows up in~$H^2$.

\section{Model Equations}\label{Section:Models}

In this section we present a simple method to generate nonlinear wave
equations with analytically known solutions, which we state explicitly
in terms of partial waves. We then list the specific models used
throughout the article. Some, but not all, of our models follow this
procedure directly.

\paragraph*{\textbf{The wave equation and partial wave solutions.}}
Let~$(r,\theta^A)$ be spherical polar coordinates built from~$x^i$ in
the usual manner. In these coordinates the flat-space wave equation
is,
\begin{align} 
\square \varphi \equiv -\p_t^2 \varphi+ \p_r^2\varphi +
\tfrac{2}{r} \p_r \varphi + \sDelta\varphi = 0\,,\label{eqn:Flat_WE}
\end{align}
with~$\sDelta$ the standard Laplacian on the round two-sphere of area
radius~$r$. The general solution can be written in terms of partial
waves~$\varphi_{lm}(t,r)$, with the full solution constructed as
\begin{align}
\varphi= \sum_{l=0}^{\infty}\sum_{m=-l}^{l}\varphi_{lm}(t,r)
Y_{lm}(\theta^A)\,,\label{eqn:partial_wave}
\end{align}
with~$Y_{lm}$ the standard spherical harmonics. Each partial wave
solves the associated equation,
\begin{align}
  -\p_t^2\varphi_{lm}+\p_r^2\varphi_{lm}+\tfrac{2}{r}\p_r\varphi_{lm}
  -\tfrac{l(l+1)}{r^2}\varphi_{lm}=0\,.
\end{align}
For our needs a convenient representation for the exact regular
solution of this equation is~\cite{GunPriPul93},
\begin{align}
\varphi_{lm}=\sum_{k=0}^l\frac{(k+l)!}{2^kk!(l-k)!}\frac{1}{r^{k+1}}
      [F^{l-k}(u)-(-1)^{l-k}F^{l-k}(v)]\,,\label{eqn:d'Alembert_soln}
\end{align}
with retarded time~$u=t-r$ and advanced time~$v=t+r$ defined in the
usual way, and~$F$ a real-valued function which we take to decay at
large argument, and which is determined by the desired initial data
for the partial wave and its time derivative.

\paragraph*{\textbf{Deformation-functions.}} To generate nonlinear
equations, we use the deformed scalar field~$\phi \equiv D(\varphi)$,
which, whenever~$\varphi$ satisfies~\eqref{eqn:Flat_WE}, must solve
\begin{align} \label{eqn:non-linear_eq}
  \square \phi-\chi\nabla_a \phi \nabla^a \phi =0\,,
\end{align}
where~$\nabla_a \phi \nabla^a \phi \equiv -(\p_t \phi)^2+ (\p_r
\phi)^2+\snabla_a\phi\snabla^a\phi$ and~$\snabla$ denotes the
covariant derivative induced by~$\eta_{ab}$ on the two-spheres of
constant~$u$ and~$v$. The deformation function~$D$ is taken to be
twice continuously differentiable and such that
\begin{align}
\chi=\tfrac{D''}{D'^2}
\end{align}
is single-valued when viewed as a function of~$\phi$. We require
moreover that~$D(\varphi)\simeq\varphi$ for small~$\varphi$. This
implies, by construction, that the model has global solutions for
small initial data, that analytic solutions can be trivially
constructed using~\eqref{eqn:d'Alembert_soln}. Moreover the manner of
blow up for larger data, should that occur, is determined by the
specific choice of~$D$. We will see below that when the deformation
function involves a periodic function this construction has to be
adjusted slightly, but the core idea is unaltered. Below we list the
models studied in the article.

\paragraph*{\textbf{Model~$\boldsymbol{1}$.}} This model is generated by
the deformation function
\begin{align} \label{model1_D}
\phi=D(\varphi)\equiv A_1^{-1}\log(1+ A_1\varphi)\,,
\end{align}
which results in the nonlinear wave equation 
\begin{align} \label{model1_EQ}
	\square \phi+A_1 \nabla_a \phi \nabla^a \phi = 0\,.
\end{align}
The parameter~$A_1$ is a real constant that we are free to
choose. Similar parameters appear in the subsequent models. This is
the classical example of Nirenberg which motivates the classical null
condition for nonlinear wave equations and was discussed
in~\cite{Kla80}. We use it primarily to determine reliability of our
code in preparation for solving models that do not arise as
deformations of the wave equation.

\paragraph*{\textbf{Model~$\boldsymbol{2}$.}} Ultimately we are interested
not in equations that arise by manipulation of the wave equation, but
those that appear in physical applications in GR. To build at least
some confidence that the properties of threshold solutions of the
former are not peculiar to those specific models, we will compute
numerical solutions for systems that cannot be constructed in the same
way. The first of these is a modification extension of model~$1$ to a
system of two coupled scalar fields. It is described by the system
\begin{align}
  &\square \phi_1+A_2 \nabla_a \phi_2 \nabla^a \phi_2 =0\,,
  \nonumber\\
  &\square \phi_2+B_2 \nabla_a \phi_1 \nabla^a \phi_1 =0\,.
  \label{model2_EQ}
\end{align}
Here we do not know solutions analytically, but in the special case
that~$A_2=B_2=A_1$ they coincide with those of~\eqref{model1_EQ}
provided that~$\phi_1$ and~$\phi_2$ and their time derivatives agree
as functions.

\paragraph*{\textbf{Model~$\boldsymbol{3}$.}} Looking at plots of the
Choptuik solution, for example Figs.~$3$ and~$4$ of~\cite{Bau18}, one
is starkly reminded of the topologists sine curve. We therefore want
to consider deformations involving periodic functions. To avoid
subtleties with branch-cuts however we adjust the construction made
in~\eqref{eqn:non-linear_eq} as follows,
\begin{align} \label{model3_D}
  &\phi_1= D_1(\varphi)\equiv A_3\sin\left[A_3^{-1}
   \log(1+ \varphi)\right] \nonumber\,, \\
  &\phi_2 = D_2(\varphi)\equiv A_3\cos\left[A_3^{-1}
   \log(1+ \varphi)\right]\,.
\end{align}
Although these deformations are not globally
invertible,~$D_1''/D_1'^2$ and~$D_2''/D_2'^2$ are single-valued
functions of both~$\phi_1$ and~$\phi_2$. Together these generate the
nonlinear coupled equations
\begin{align} \label{model3_EQ}
  &\square \phi_1+\frac{\phi_1+A_3 \phi_2}{A_3^2-\phi_1^2}
  \nabla_a \phi_1 \nabla^a \phi_1 =0 \nonumber\,, \\
  &\square \phi_2+\frac{\phi_2-A_3 \phi_1}{A_3^2-\phi_2^2}
  \nabla_a \phi_2 \nabla^a \phi_2 =0 \,,
\end{align}
with the algebraic constraint~$\phi_1^2+\phi_2^2=A_3^2$. Using the
constraint we obtain
\begin{align}
  \frac{\nabla_a \phi_1 \nabla^a \phi_1}{A_3^2-\phi_2^2}
  -\frac{\nabla_a \phi_2 \nabla^a \phi_2}{A_3^2-\phi_1^2}=0\,,
\end{align}
and also
\begin{align}
  \phi_1 \square \phi_1+\phi_2 \square \phi_2
  + \nabla_a \phi_1 \nabla^a \phi_1+\nabla_a \phi_2
  \nabla^a \phi_2 &=0\,,
\end{align}
which, with system~\eqref{model3_EQ}, results in,
\begin{align}
  \nabla_a \phi_1 \nabla^a \phi_1+\nabla_a \phi_2
  \nabla^a \phi_2=A_3^2 \frac{\nabla_a
    \phi_1 \nabla^a \phi_1}{A_3^2-\phi_2^2}.
\end{align}
Using this relation it is easy to see that system~\eqref{model3_EQ},
subject to the algebraic constraint, is equivalent to
\begin{align} \label{model3_EQ_reg}
  &\square \phi_1+A_3^{-2}\left(\phi_1+A_3 \phi_2\right)
  \left(\nabla_a \phi_1 \nabla^a \phi_1+\nabla_a \phi_2
  \nabla^a \phi_2\right) =0 \nonumber\,, \\
  &\square \phi_2+A_3^{-2}\left(\phi_2-A_3 \phi_1\right)
  \left(\nabla_a \phi_1 \nabla^a \phi_1+\nabla_a
  \phi_2 \nabla^a \phi_2\right) =0 \,.
\end{align}
For the Cauchy problem, solutions with initial data satisfying the
constraint will be of the form~\eqref{model3_D}, and thus satisfy the
constraint everywhere for all times. For the initial boundary value
problem boundary conditions must be constraint preserving. At the
continuum level there is thus no clear advantage
of~\eqref{model3_EQ_reg} over~\eqref{model3_EQ}, but crucially for
numerical approximation we avoid the explicit poles present in the
latter. By using the constraint to eradicate either~$\phi_1$
or~$\phi_2$ in Eqn.~\eqref{model3_EQ}, we see that the fields satisfy
equations similar to~\eqref{eqn:dss_model_simple}.

\paragraph*{\textbf{Model~$\boldsymbol{4}$.}} Just as we view model~$2$
as an extension model~$1$, in model~$4$ we extend model~$3$ by
dropping the algebraic constraint on~$\phi_1^2+\phi_2^2$. We
simultaneously adjust the equations of motion to
\begin{align}
  \label{model4_EQ}
  &\square \phi_1+A_4^{-2}\left(\phi_1+A_4 \phi_2\right)
  \left(\nabla_a \phi_1 \nabla^a \phi_1+\nabla_a \phi_2
  \nabla^a \phi_2\right) =0 \nonumber\,, \\
  &\square\phi_2+B_4^{-2}\left(\phi_2-B_4 \phi_1\right)
  \left(\nabla_a \phi_1 \nabla^a \phi_1
  +\nabla_a \phi_2\nabla^a \phi_2\right) =0\,.
\end{align}
Again, since this model was not obtained from a deformation of the
wave equation, solutions of this system are not known analytically in
general, but are coincident when the constraint is satisfied
and~$A_4=B_4$. Blow-up solutions will be investigated carefully in the
section~\ref{Section:Critical}, but it is already obvious that blow-up
solutions for model~$3$ will be oscillatory in nature. The key
question here, which we examine numerically in
section~\ref{Section:Numerics}, is whether or not this behavior
persists generically with the present model.

\paragraph*{\textbf{Model~$\boldsymbol{5}$.}} Returning to the general
deformation function, we can define the conformal
metric~$\tilde{\eta}_{ab}=\Omega^2\eta_{ab}$ with conformal
factor~$\Omega^{-2}=\p_{\varphi}D$ viewed now as a function of~$\phi$.
We denote the inverse conformal metric by~$\tilde{\eta}^{ab}$ and the
associated covariant derivative by~$\tilde{\nabla}_a$. In these terms
the general deformation equation~\eqref{eqn:non-linear_eq} can be
rewritten as,
\begin{align}
\tilde{\Box}\phi\equiv\tilde{\eta}^{ab}\tilde{\nabla}_a\tilde{\nabla}_b\phi
=0\,.\label{eqn:conformal_wave}
\end{align}
It follows immediately that the deformed wave equation admits the
standard stress-energy,
\begin{align}
  T_{ab}[\phi]&=\tilde{\nabla}_a\phi\tilde{\nabla}_b\phi
  -\tfrac{1}{2}\tilde{\eta}_{ab}\tilde{\nabla}^c\phi\tilde{\nabla}_c\phi
  \,,\label{eqn:conformal_SE}
\end{align}
where, as is conventional, indices on the conformal covariant
derivatives were raised using the conformal metric. The
stress-energy~\eqref{eqn:conformal_SE} is of course covariantly
conserved~$\tilde{\nabla}^bT_{ab}[\phi]=0$. In this manner we have
rewritten our original model in a clean geometric form that resulted
in a quasilinear equation. This reformulation by itself is not
particularly helpful, but in the future the conserved energy will
certainly be useful in proving the findings of this paper rigorously.
This construction also allows us to build more general `DSS' type
models too. Let~$\phi=\phi_0$, and suppose that the original
deformation function~$D$ is monotonic increasing on its
domain. Inspired by~\eqref{model3_D}, set
\begin{align}
  \phi_1&= P_1(\phi_0) \,,\qquad
  \phi_2= P_2(\phi_0) \,,
  \label{eqn:model5_periodic_D}
\end{align}
where~$P_1$ and~$P_2$ are any periodic functions that satisfy
\begin{align}
  P_1^2+P_2^2=\delta^2\,,\qquad P_1'^2+P_2'^2=\varepsilon^2\,,
\end{align}
with~$\delta$ and~$\varepsilon$ positive functions of~$\phi_0$
uniformly bounded above and below away from~$0$. Computations very
similar to those for model~$3$ in the build up
to~\eqref{model3_EQ_reg} then reveal the regularized equations of
motion,
\begin{align}
  \tilde{\Box}\phi_1
  -\varepsilon^{-2}P_1''(\phi_0)
  \big(
  \tilde{\nabla}^a\phi_1\tilde{\nabla}_a\phi_1
  +\tilde{\nabla}^a\phi_2\tilde{\nabla}_a\phi_2
  \big)=0\,,\nonumber\\
  \tilde{\Box}\phi_2
  -\varepsilon^{-2}P_2''(\phi_0)
  \big(
  \tilde{\nabla}^a\phi_1\tilde{\nabla}_a\phi_1
  +\tilde{\nabla}^a\phi_2\tilde{\nabla}_a\phi_2
  \big)=0\,.\label{eqn:model5_periodic_EOMs}
\end{align}
These equations can be solved alongside~\eqref{eqn:conformal_wave} for
a complete model. The fields~$\phi_1$ and~$\phi_2$ have a combined
conserved stress-energy that can again be obtained naturally by a
conformal transformation. This model has the disadvantage of requiring
more fields, but is more robust than model~$3$, because it grants a
large amount of freedom in choosing a compactifying function. A
shortcoming of using~\eqref{eqn:conformal_wave}
with~\eqref{eqn:model5_periodic_EOMs} is that the coupling between the
fields is one-directional, which makes it impossible, when choosing
initial data, that violate the various constraints between the
different fields, to seed non-trivial evolution in~$\phi_0$
from~$\phi_1$ and~$\phi_2$. A final modification can be made to
sidestep this. Using
\begin{align}
  \varepsilon^2\tilde{\nabla}^a\phi_0\tilde{\nabla}_a\phi_0=
  \tilde{\nabla}^a\phi_1\tilde{\nabla}_a\phi_1
  +\tilde{\nabla}^a\phi_2\tilde{\nabla}_a\phi_2\,,
\end{align}
Eqn.~\eqref{eqn:conformal_wave} can be rewritten as,
\begin{align}
  \hat{\Box}\phi_0
  -\varepsilon^{-2}\chi
  \big(\tilde{\nabla}^a\phi_1\tilde{\nabla}_a\phi_1
  +\tilde{\nabla}^a\phi_2\tilde{\nabla}_a\phi_2
  \big)=0\,,\label{eqn:conformal_reduced}
\end{align}
where~$\hat{\Box}\phi_0\equiv\tilde{\eta}^{ab}\nabla_a\nabla_b\phi_0$
denotes the reduced wave operator associated with~$\tilde{\eta}_{ab}$,
and~$\chi$ is again to be viewed as a function
of~$\phi_0$. Interestingly, the combined
system~\eqref{eqn:model5_periodic_EOMs},\eqref{eqn:conformal_reduced}
admits a natural analogy with GR. The fields~$\phi_1,\phi_2$ are akin
to some field theory matter and, since it is required in
building~$\tilde{\eta}_{ab}$, the field~$\phi_0$ to a metric
component.

\section{Criticality, regularity and the threshold of blow-up}
\label{Section:Critical}

In this section we focus on nonlinear equations, as exemplified by
models~$1$ and~$3$, that are generated by a deformation of the wave
equation. We examine the extent to which threshold solutions and those
in a neighborhood of the threshold in phase space exhibit a behavior
like that in gravitational collapse. We start with spherical solutions
and then move on to the more general setting.

\paragraph*{\textbf{Bounds and blow-up in spherical symmetry.}} We want to
establish that spherical threshold solutions blow up at the origin. We
start with solutions to the wave equation. In this context the
d'Alembert solution~\eqref{eqn:d'Alembert_soln} takes the well-known
form
\begin{align}
  \varphi&=\frac{1}{r}\,[F(t+r)-F(t-r)]\,.
  \label{eqn:d'Alembert_sph_soln}
\end{align}
Consider a subset~$\{\varphi_{\star}(t,r)\}$ of the
solutions~\eqref{eqn:d'Alembert_sph_soln}, such that
for~$t<t_{\star}$, we have~$\varphi_{\star}(t,r)>\xi_{\star}$, some
constant, and~$\varphi_{\star}(t_{\star},r_{\star})=\xi_{\star}$ is a
local minimum. Loosely speaking we may think of the
point~$(t_{\star},r_{\star})$ as the location of blow-up in the
deformed equation, so that the label~$\star$, somewhat prejudicially,
stands for ``critical''.  This minimum must be attained at the
origin,~$r_{\star}=0$. To see this, suppose on the contrary
that~$r_{\star}>0$. Since~$(t_{\star},r_{\star})$ is a local extremum
we have,
\begin{align}
  r_{\star}\left[F_\star'(t_{\star}+r_{\star})+F_\star'(t_{\star}-r_{\star})\right]
  &=F_\star(t_{\star}+r_{\star})-F_\star(t_{\star}-r_{\star})\,,\nonumber\\
  F_\star'(t_{\star}+r_{\star})&=F_\star'(t_{\star}-r_{\star})\,,
\end{align}
which implies that
\begin{align}
\varphi_{\star}(t_{\star},r_{\star})=2F_\star'(t_{\star}-r_{\star})=\xi_{\star}\,.
\end{align}
At the origin however we have
\begin{align}
\varphi_{\star}(t,0)=2F_\star'(t)\,,
\end{align}
which gives
\begin{align}
\varphi_{\star}(t_{\star}-r_{\star},0)=\varphi_{\star}(t_{\star},r_{\star})=\xi_{\star}\,.
\end{align}
By assumption~$r_{\star}>0$, so this contradicts the assumption
that~$\varphi_{\star}(t,0)>\xi_{\star}$ for~$t<t_{\star}$. Thus we
have shown that~$r_{\star}=0$. Consequently the global minimum of a
spherical solution to the wave equation occurs at the origin. Consider
now a compactifying deformation function~$D\left[\varphi\right]=
\mathcal{C}\left(\varphi\right)$, with~$\mathcal{C}(\varphi)$ defined
on~$\varphi>\xi_{\star}$ and such that we have the blow-up
\begin{align}
\lim_{\varphi \to \xi_{\star}} \mathcal{C}(\varphi)=\infty.
\end{align}
Recall from Section~\ref{Section:Models} that we additionally
require~$\mathcal{C}(\varphi)\simeq\varphi$ for small~$\varphi$. For a
one-parameter family of initial data, the solutions of
Eqn.~\eqref{eqn:non-linear_eq} at the threshold between global
existence and blow-up in are of the
form~$\phi_{\star}(t,r)\equiv\mathcal{C}\left[\varphi_{\star}(t,r)\right]$. These
are called threshold solutions, and by the previous discussion must
blow up at~$(t_{\star},0)$.

\paragraph*{\textbf{Criticality of spherical threshold solutions.}}
Interestingly, the threshold solutions of our deformation models are
universal in the sense that the form of their blow-up
near~$(t_{\star},0)$ is independent of the initial conditions and,
thus, of the family of initial data considered. We therefore call this
``late time'' universal solution a \textit{critical solution}. To
illustrate this notice that the original solution to the wave equation
satisfies,
\begin{align}
  &\lim_{(t,r)\to(t_{\star},0)}\varphi_{\star}(t,r) \sim \xi_{\star}
  +\tfrac{1}{2}\p_t^2\varphi_{\star}(t_{\star},0)
  \left(t_{\star}-t\right)^2\nonumber\\
  &\qquad-\tfrac{1}{2}\p_t \p_r\varphi_{\star}(t_{\star},0)
  \left(t_{\star}-t\right) r+\tfrac{1}{2}\p_r^2
  \varphi_{\star}(t_{\star},0)r^2\,.
\end{align}
Moreover, it is easy to show that~$\p_t \p_r\varphi(t,0)=0$
and~$\p^2_t\varphi(t,0)=3\p^2_r\varphi(t,0)=2 F'''(t)$ for any regular
solution~\eqref{eqn:d'Alembert_soln} of the wave equation. The last
limit thus becomes 
\begin{align}
\lim_{(t,r)\to(t_{\star},0)} \varphi_{\star}&(t,r)\sim
\xi_{\star}+2F_{\star}'''(t_{\star})\left[\left(t_{\star}-t\right)^2
  +\tfrac{1}{3}r^2\right]\nonumber\\
&\sim\xi_{\star}+2F_{\star}'''(t_{\star})e^{-2T}\left(1+\frac{1}{3}X^2\right)\,,
\end{align}
where in the last line we have introduced similarity adapted
coordinates
\begin{align}
T=-\log(t_\star-t)\,,\qquad X=(t_\star-t)^{-1}r\,,
\end{align}
and expanded about~$(t_\star,r_\star)$. Working with model~$1$ and
setting~$A_1=1$ we
have~$\mathcal{C}(\varphi)=\log\left(1+\varphi\right)$. Then~$\xi_{\star}=-1$,
which gives
\begin{align}
  &\lim_{(t,r)\to(t_{\star},0)} \phi_{\star}(t,r) \sim \nonumber\\
  &\sim-2T+\log \left(1+\tfrac{1}{3} X^2\right)+\log
  \left[2F_\star'''(t_\star)\right] \,,
\end{align}
where the first term is the critical solution; note that in the
neighborhood of~$(t_\star,r_\star)$, within its past light-cone, we
have~$X\leq1$. To leading order this expression is independent of the
initial data, which illustrates the universality of blow-up of
threshold solutions. Evidently the critical solution blows up
in~$L^\infty$. Regularity in other function spaces is discussed
below. The critical solution is \textit{approximately} CSS, centered
at the blow-up point, with~$\nu=0$ (see Eqn.~\eqref{self-sim}),
\begin{align}
\lim_{(t',r)\to(0,0)} \phi_{\star}(t_{\star}+\lambda t',\lambda r) \sim
\phi_{\star}(t_{\star}+ t', r)\,.
\end{align}

\paragraph*{\textbf{Alternative compactifications.}} For a more general
class of models with~$\xi_{\star}=-1$, we consider
\begin{align}
  \mathcal{C}(\varphi)=\frac{1}{n}\Big(1-\frac{1}
          {(1+\varphi)^n}\Big),\label{eqn:compactification}
\end{align}
where~$n>0$, and so one has
\begin{align}
\lim_{(t,r)\to(t_{\star},0)} \phi_{\star}(t,r) \sim 
 \tfrac{1}{n}
\left[2F_{\star}'''(t_{\star})\right]^{-n}\left(1+\tfrac{1}{3}X^2\right)^{-n}e^{2nT}\,.
\end{align}
In this case, the universality of blow-up of threshold solutions is
weaker since there is a dependence on the initial conditions
through~$\p_t^2\varphi_{\star}(t_{\star},0)$. Nevertheless we still
have a universal power~$2n$. It is remarkable that the entire freedom
within a large function space boils down to just one parameter at the
threshold. It is appealing to think of the single remaining parameter
as a single hair of `the' critical solution, so that uniqueness can be
understood in a parameterized sense as in the standard discussion of
stationary black holes with symmetry.  The threshold solutions of
these models blow up in a CSS manner, centered at the blow-up point,
with~$\nu=-2 n$ (see Eqn.~\eqref{self-sim}). 

\paragraph*{\textbf{Deformations using periodic functions.}} Now let us
focus on a deformation with the functional form~$D\left[\varphi\right]
\equiv\mathcal{P}\circ \mathcal{C}(\varphi)$, with~$\mathcal{P}$ a
bounded periodic function with period~$\Lambda$,
satisfying~$\lim_{\mathcal{C}\to0}\mathcal{P}(\mathcal{C})\sim\mathcal{C}$. By
construction, the solutions of Eqn.~\eqref{eqn:non-linear_eq} have
global existence for sufficiently small initial data and can never
blow up in~$L^\infty$ regardless of the initial conditions. First
derivatives of solutions with sufficiently large initial data,
however, must explode. Here the threshold solutions are the ones at
the threshold between global existence and this blow-up, and are of
the form~$\phi_{\star}(t,r)\equiv \mathcal{P}\circ
\mathcal{C}\left[\varphi_{\star}(t,r)\right]$. Similarly to the
previous type of deformation functions, the blow-up of these threshold
solutions is universal and happens at~$(t_{\star},0)$. For this type
of deformation function we have the first derivatives
\begin{align}
  \p_t\phi_{\star}(t,r) &= \mathcal{P}'\circ \mathcal{C}
  \left[\varphi_{\star}(t,r)\right]
  \mathcal{C}'\left[\varphi_{\star}(t,r)\right]
  \p_t \varphi_{\star}(t,r)\,, \nonumber\\
  \p_r\phi_{\star}(t,r) &= \mathcal{P}'\circ \mathcal{C}
  \left[\varphi_{\star}(t,r)\right]
  \mathcal{C}'\left[\varphi_{\star}(t,r)\right]
  \p_r \varphi_{\star}(t,r)\,.
\end{align}
Model~$3$ has~$\xi_{\star}=-1$, period~$\Lambda=2\pi$ and
\begin{align}
  \phi_1=\mathcal{P} \circ
  \mathcal{C}(\varphi)=
  A_3\sin\left[A_3^{-1}\log(1+\varphi)\right]\,,
\end{align}
which results in the bounded field
\begin{align}
  &\lim_{(t,r)\to(t_{\star},0)}  \phi_{1\star}(t,r) \sim \nonumber \\
  &A_3\sin\left[A_3^{-1}\log\left(2F_{\star}'''(t_{\star})e^{-2T}
    \left[1+\frac{1}{3}X^2\right]
  \right)\right]\,,
\label{eqn:SinLogCritical}
\end{align}
and the blow-up of the first derivatives
\begin{align}
  &\lim_{(t,r)\to(t_{\star},0)} \p_t \phi_{1\star}(t,r)
  \sim -\frac{2e^{T}}{1+\tfrac{1}{3}X^2}\cos(\star)\,,
\end{align}
and
\begin{align}
  &\lim_{(t,r)\to(t_{\star},0)} \p_r\phi_{1\star}(t,r)
  \sim \frac{2 Xe^{T}}{3+X^2}\cos(\star)\,.
\end{align}
where~$\star$ here denotes the argument of the~$\sin$ term
in~\eqref{eqn:SinLogCritical}. Thus the threshold solutions of this
model blow up, and there are universal powers directly prior. Again,
dependence on initial data reduces down to just one number, in this
case appearing as a pure phase off-set. An interesting challenge for
either this model or any other would be to diagnose such behavior by
purely numerical means. The blow-up of~$\p_t \phi_{1 \star}$
and~$\p_r\phi_{1 \star}$ is DSS, centered at~$(t_{\star},0)$,
with~$\nu=-1$ and~$\lambda_m=e^{-m \Delta}=e^{ m \pi A_3}$ (see
Eqn.\eqref{self-sim}),
\begin{align}
  \lim_{(t',r)\to(0,0)} \p_\mu \phi_{1 \star}(t_{\star}+\lambda_n t',\lambda_n r)
  \sim \lambda_n^{-1} \p_\mu \phi_{1 \star}(t_{\star}+ t', r)\,.
\end{align}
Using the construction of model~$5$ we can build
alternative~$\mathcal{P}\circ\mathcal{C}$ deformation models.  For
example, by combining the
compactification~\eqref{eqn:compactification} with~$\sin$, we get
\begin{align} \label{eqn:SinCCritical}
  \phi_{1}=\sin\left[\frac{1}{n}
  \left(1-\frac{1}{(1+\varphi)^n}\right)\right]\,.
\end{align}
The threshold solutions of this model have the form
\begin{align}
&\lim_{(t,r)\to(t_{\star},0)}  \phi_{1\star}(t,r) \sim \nonumber \\
  &\quad \sin \left(\tfrac{1}{n}
  \left[2F_{\star}'''(t_{\star})\right]^{-n}
  \left[\left(t_{\star}-t\right)^2
    +\tfrac{1}{3}r^2\right]^{-n} \right)\,,
\end{align}
which is bounded. Their first derivatives blow up with
\begin{align}
  &\lim_{(t,r)\to(t_{\star},0)} \p_t \phi_{1\star}(t,r) \sim \nonumber\\
  &\quad 2\left[2F_{\star}'''(t_{\star})\right]^{-n}
  \frac{t_\star-t}{\left(\left[t_{\star}-t\right]^2
  	+\tfrac{1}{3}r^2\right)^{n+1}}\cos(\star)\,,
\end{align}
and
\begin{align}
&\lim_{(t,r)\to(t_{\star},0)} \p_r \phi_{1\star}(t,r) \sim \nonumber\\
  &\quad -\tfrac{2}{3}\left[2F_{\star}'''(t_{\star})\right]^{-n}
  \frac{r}{\left(\left[t_{\star}-t\right]^2
	+\tfrac{1}{3}r^2\right)^{n+1}}\cos(\star)\,,
\end{align}
where~$\star$ here denotes the argument of the~$\sin$ term
in~\eqref{eqn:SinCCritical}. It is easy to verify (looking at
the~$\cos$ term) that in these coordinates the blow-up does not
satisfy the symmetry~\eqref{self-sim}. We have not found a coordinate
system, which would imply a DSS blow-up, in which that property holds;
however, this possibility is not excluded. Nevertheless the power of
blow-up is still universal and (as before) it is~$2n$. Again, the
critical solution is unique modulo a single parameter. It is very
interesting that much of the desired phenomenology can be achieved but
with threshold solutions of an apparently different character. If we
insisted on finding alternative models that do have DSS threshold
solutions we could try deformation functions of the form,
\begin{align}
  D(\varphi)\simeq e^{\mathcal{C}(\varphi)}
  \mathcal{P}\circ\mathcal{C}(\varphi)\,.
\end{align}
but we are already content with the simpler option above. All of the
power-laws discussed so far appear in physical space. Below we discuss
similar results in phase space~($a-a_\star$).

\begin{figure*}[t] 
\centering
\includegraphics[width=1.0\textwidth]{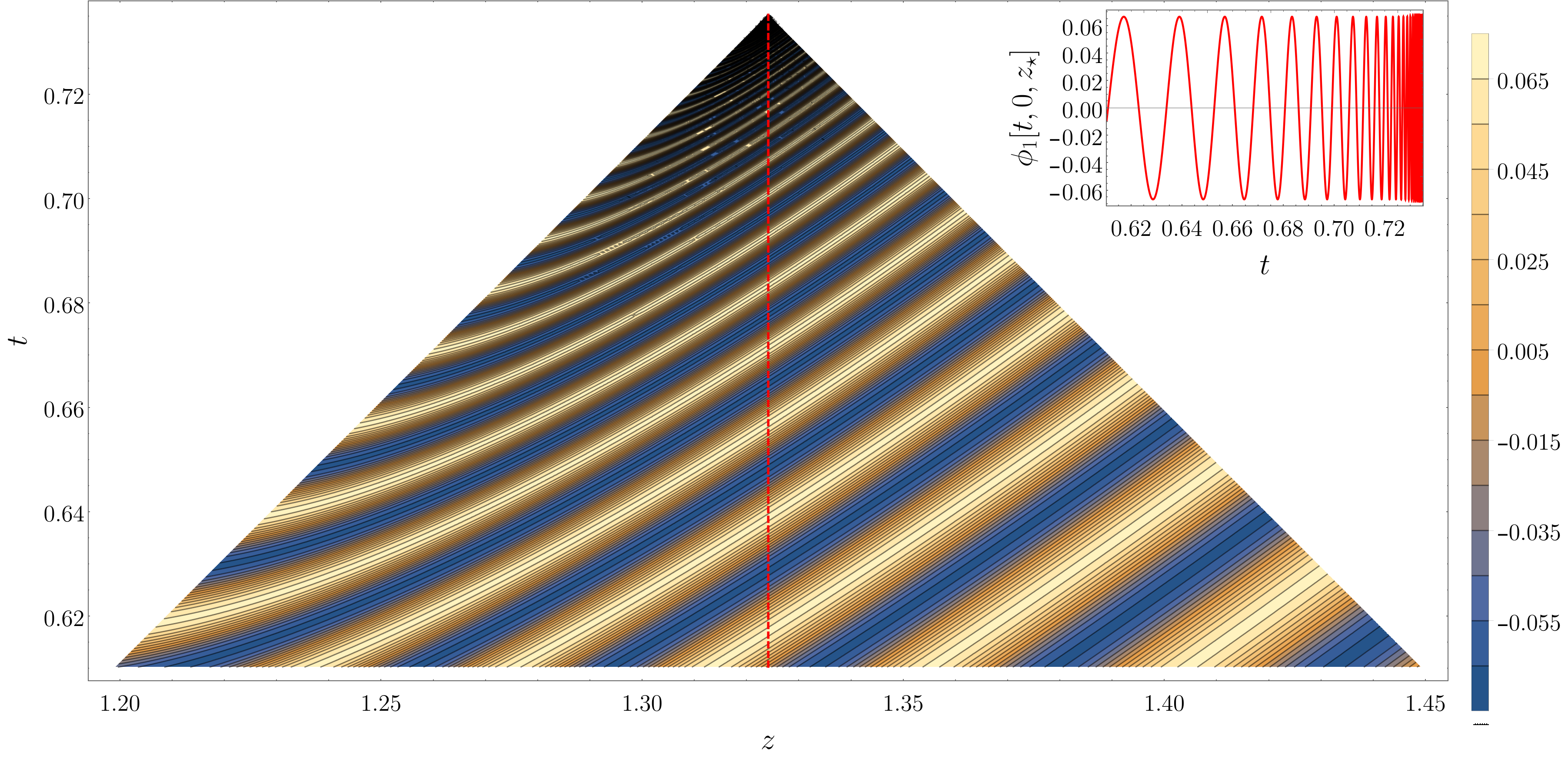}
\caption{A contour plot of an axisymmetric threshold solution for
  model~$3$ shown on the symmetry axis. Despite shared attributes with
  the spherical solution of Fig.~\ref{Fig:Critical}, there are obvious
  differences too, as the data here leading to blow-up is mostly
  outgoing. For this model therefore the conjecture that there is in
  general a unique threshold solution regardless of initial data is
  false.}\label{Fig:l2DSS}
\end{figure*}

\begin{figure}[t]
\centering
\includegraphics[width=0.48\textwidth]{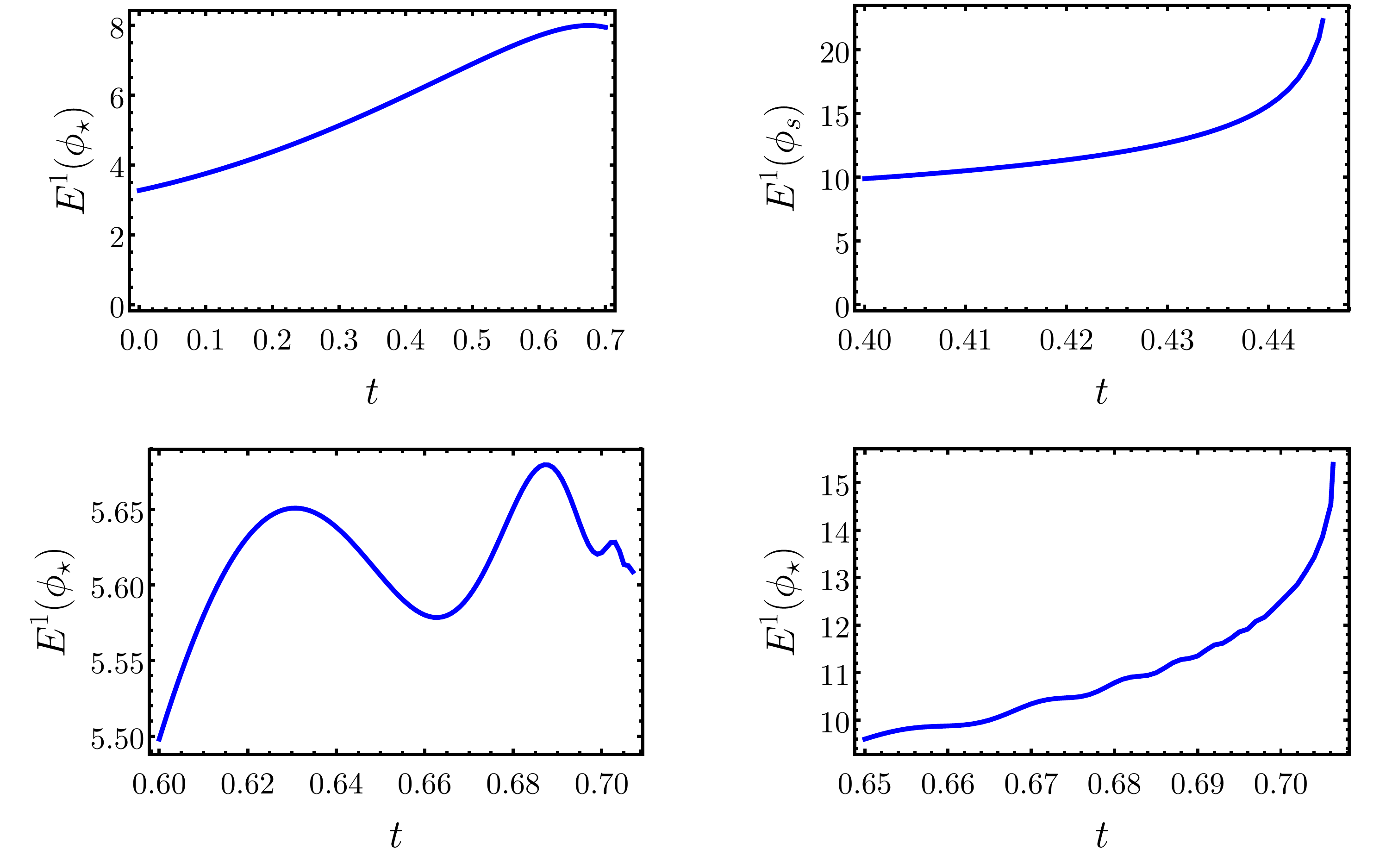}
\caption{Plots of the~$E^1$ norm for spherical solutions of various
  models up to the time at which some field quantity explodes
  in~$L^\infty$. On the top left we have a threshold solution for
  model~$1$. On the top right a supercritical solution for the same
  model is shown, demonstrating that a variation of behavior is
  possible at blow-up. On the bottom left we show the result
  for~$\phi_1$ from model~$3$ at the threshold. Finally in the lower
  right panel we show the same for the composite deformation
  function~$\sin\circ\,\mathcal{C}$, with the
  compactification~\eqref{eqn:compactification} and~$n=1/4$, which can
  be used in practice within model~$5$. These examples are compatible
  with our consideration of self-similar functions and our naive norm
  estimates.\label{Fig:NormPlot}}
\end{figure}

\paragraph*{\textbf{Regularity of spherical solutions at blow-up.}} So
far, we have focused only on pointwise blow-up, but a proper
understanding of the threshold must also include statements about
local integrability. Consider first deformation functions that involve
only a compactification. As we have already discussed, with this setup
blow-up solutions, whether generic or at the threshold, become
unbounded pointwise. Therefore by Sobolev embedding~$H^2$ must explode
(see Eqn.~\eqref{SobolevEmbed}), but beneath that, the story is more
subtle.  By choosing initial data constant in space for the first time
derivative out to some radius and then cutting off, it is clear that
solutions can blow up in~$L^2$ for any of our pure compactification
deformation functions. But around the threshold, the solutions blow up
in a special, localized manner, so that boundedness in~$L^2$ depends
on the specific deformation function / compactification. This must
also fall-in line with the observations made in the previous section
about regularity of self-similar functions. In fact, since the
compactification determines also~$\nu$, there must exist a
relationship between the universal powers and regularity at the
threshold. To examine this, we suppose that the integral is dominated
by the values of the integrand at the origin. Expanding then, we find
with the log compactification that
\begin{align}
  ||\phi_\star||_{L^2}&\sim e^{-3T/2}T\,,&\quad ||\phi_\star||_{E^1}
  &\sim e^{-T/2}\,,
  \nonumber\\
  ||\phi_s||_ {L^2}&\sim T\,,&\quad ||\phi_s||_{E^1}&\sim e^{T}\,,
\end{align}
for threshold solutions. Here we used the fact that, at the threshold,
the spatial scale on which the solution becomes large pointwise is
fixed in the similarity coordinate~$X$. We assumed that blow-up of the
supercritical solution~$\phi_s$ occurred at the origin with the
spatial scale fixed in~$r$, and set the
slow-time~$T=-\log(t_\star-t)$, with~$t_\star$ the instant at which
the solution explodes so that~$T\to\infty$ at the blow-up. Thus this
estimate on~$\phi_s$ need not be verified in practice, and indeed it
is easy to come up with examples in which~$L^2(\phi_s)$ is finite even
at blow-up. For the alternative
compactification~\eqref{eqn:compactification} we find
\begin{align}
  ||\phi_\star||_{L^2}&\sim e^{(2n-3/2)T}\,,
  &\quad ||\phi_\star||_{E^1}&\sim e^{(2n-1/2)T}\,,\nonumber\\
  ||\phi_s||_{L^2}&\sim e^{nT}\,,&\quad ||\phi_s||_{E^1}&\sim e^{(n+1)T}\,.
\end{align}
Again these naive estimates on~$\phi_s$ need not be satisfied, and
serve only as an indication of possible behavior. All of these
estimates can be verified numerically and are in agreement with the
results in Section~\ref{Section:Self-similar}. Moving on to
deformation functions involving a periodic function, by construction,
obviously solutions can never blow up in~$L^2$. Proceeding as before,
we have
\begin{align}
  ||\phi_\star||_{E^1}&\sim e^{-T/2}\,,\quad &||\phi_s||_{E^1}\sim e^T\,,
\end{align}
for model~$3$ and
\begin{align}
  ||\phi_\star||_{E^1}&\sim e^{(2n-1/2)T}\,,\quad &||\phi_s||_{E^1}
  \sim e^{(n+1)T}\,,
\end{align}
with the composite deformation function~$\sin\circ\,\mathcal{C}$
taking again the compactification~\eqref{eqn:compactification}. As
mentioned in the discussion above, we have checked these predictions
in practice by computing numerically norms for different blow-up
solutions. Some examples are shown in Fig.~\ref{Fig:NormPlot}. In
summary, threshold solutions blow up at~$t=t_{\star}$ in~$H^1$
when~$n\geq1/4$ (that is~$\nu\leq-1/2$), and in the CSS setting
in~$L^2$ when~$n\geq3/4$ ($\nu\leq-3/2$). The two take-aways are
first, that generic blow-up solutions are less regular than threshold
solutions, and second that there is a direct relationship between the
universal power and the specific level of regularity.

\paragraph*{\textbf{Aspherical perturbations of spherical critical solutions.}}
So far we have established that in pure spherical symmetry threshold
solutions of our deformation models depend to leading order on only
one number from the initial data and are, in this sense,
unique. Therefore, in accord with the usual picture of critical
gravitational collapse, if we consider a one-parameter family of
spherically symmetric initial data and tune this parameter to the
threshold of blow-up we recover the critical solution. What is more,
simply by continuous dependence on given data, spherical initial data
close to the threshold generate solutions that appear like the
critical solution for some time in their development. Evidently the
latter statement is true also for nonspherical perturbations of the
spherical critical solution. But in fact a stronger result holds.
Take a family of spherical
solutions~$\phi_a(t,r)=D[a\,\varphi_\star(t,r)]$ normalized so
that~$a=1$ corresponds to the threshold
solution~$\phi_\star=\phi_1$. As discussed above, in the past light
cone of the blow-up point,~$\phi_\star$ is associated with a critical
solution by simple Taylor expansion. Let~$\tilde{\varphi}$ denote {\it
  any} regular partial wave solution~\eqref{eqn:partial_wave} with
vanishing spherical component~$\tilde{\varphi}_{00}$. We may think of
this solution as being parameterized by the infinite number of
parameters stating how much of each of the individual partial wave
solutions~$\tilde{\varphi}_{lm}$, each of which also have a full
functional degree of freedom, is present. Consider the perturbed
solutions
\begin{align}
  \tilde{\phi}_a=D[a\,(\varphi_\star+\epsilon\,\tilde{\varphi})]\,,
  \label{eqn:perturbed_spherical_critical}
\end{align}
and observe, crucially, from~\eqref{eqn:d'Alembert_soln}
that~$\tilde{\varphi}_{lm}(t,r)= O(r^l)$ near the origin. We then see
that for~$\epsilon$ sufficiently
small~$\tilde{\phi}_\star=\tilde{\phi}_1$ is {\it also} a threshold
solution. Starting from~$\tilde{\phi}_\star$, within this family the
{\it only} way to restore global existence is to reduce the strength
parameter~$a$. It seems that this result would fit nicely with a
perturbative analysis along the lines of that given
in~\cite{GarGun98}. To understand the effect of the~$\tilde{\varphi}$
on the asymptotic solution in the past light-cone of the blow-up point
we present below a generalization of the spherical Taylor expansion
given above.

\begin{figure}[t]
\centering
\includegraphics[width=0.47\textwidth]{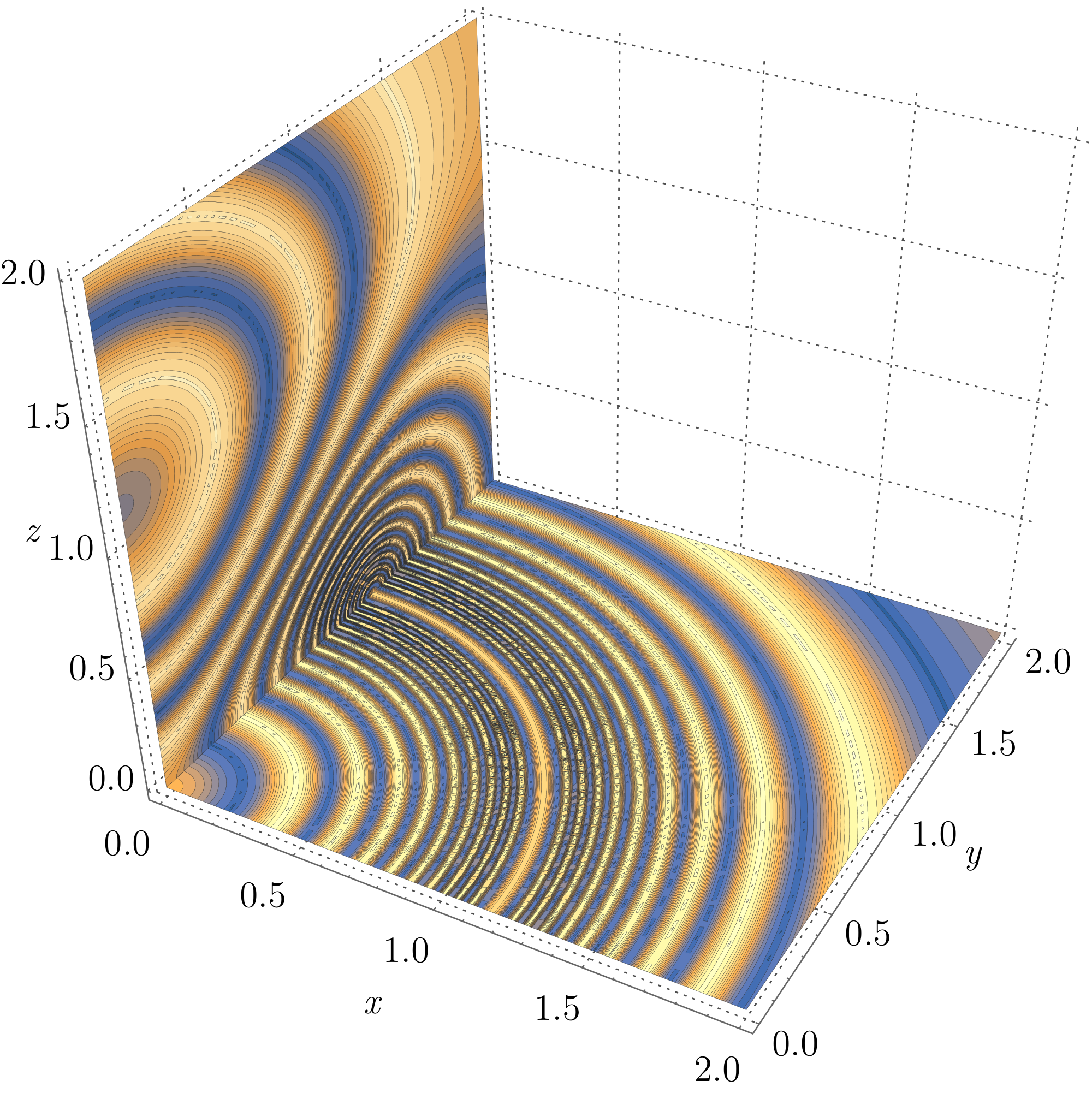}
\caption{Here we plot a pure~$l=2$, $m=0$ threshold solution for
  model~$3$ shortly before blow-up. Special in this case is that the
  blow-up occurs on a ring in the~$z=0$ plane. This was achieved with
  the family~\eqref{eqn:Ring_ID}, which we may think of as the same
  data as~\eqref{eqn:axis_ID}, but evolved backwards in time. This
  shows that away from spherical symmetry, even when building
  threshold solutions purely from a single harmonic, there exist
  fundamentally different threshold solutions, although the number of
  such branches for each harmonic is always presumably finite. This
  story becomes even more involved with higher harmonics.}
\label{Fig:Ring}
\end{figure}

\paragraph*{\textbf{Single-harmonic threshold solutions.}} To this point
the behavior exhibited by solutions of our models had a direct analog
with the standard picture of critical collapse. In moving to consider
general nonspherical threshold solutions we now part ways with that
picture. The discussion here is focused on model~$3$, but holds in
fact more generally. We start by constructing a particular threshold
solution from a pure~$l=2$, $m=0$ partial wave solution~$\varphi_{20}$
to the wave equation. Recalling the exact
solution~\eqref{eqn:d'Alembert_soln} and working with the family
generated by the Gaussian,
\begin{align}
  F(r)&=a e^{-(r+1)^2}\,,\label{eqn:axis_ID}
\end{align}
we find that the threshold solution~$\phi_{\star}$ is obtained
with~$a_\star\simeq1.678$. As observed above, the partial wave
vanishes at the origin, and therefore the blow-up point occurs
elsewhere, in this case
at~$(t_\star,x_\star,y_\star,z_\star)\simeq(0.735,0,0,1.324)$. This
threshold solution is plotted in Fig.~\ref{Fig:l2DSS}. Although there
clear are qualitative similarities with the spherical threshold
solution plotted in Fig.~\ref{Fig:Critical} for the same model one
could hardly claim that the two solutions are the same. Interestingly,
even if we restrict to threshold solutions built from a single
spherical harmonic in this manner there is still another distinct
branch of threshold solutions. To see this consider, for example, the
form of the~$Y_{20}$ harmonic,
\begin{align}
  Y_{20}&= \frac{1}{4}\sqrt{\frac{5}{\pi}}\big(3\cos^2\theta^1-1\big)\,,
\end{align}
which has local extrema on the~$x$ and~$z$-axes. Since we are
concerned here with an axisymmetric solution we are free to
identify~$x$ with the cylindrical radial coordinate. Therefore our
solution to the wave equation
\begin{align}
  \varphi&=\varphi_{20}Y_{20}
\end{align}
giving rise to a solution of the deformed wave equation can explode
the compactification in one of two ways, 
\begin{align}
  \varphi_{20}&=-2\sqrt{\frac{\pi}{5}}\,, &\quad
  \varphi_{20}&=4\sqrt{\frac{\pi}{5}} \,,
\end{align}
at some point, resulting in the first case in blow-up of~$\phi$ on the
symmetry axis as plotted in Fig.~\ref{Fig:l2DSS}, or else {\it on a
  ring} in the~$x\-y$-plane in the second. A snapshot of a solution
close to this type of blow-up, obtained with the family
\begin{align}
G(r)&=-F(-r)\,,\label{eqn:Ring_ID}
\end{align}
with~$F$ the Gaussian from before, is shown in Fig.~\ref{Fig:Ring}.

\begin{figure}[t]
\centering
\includegraphics[width=0.47\textwidth]{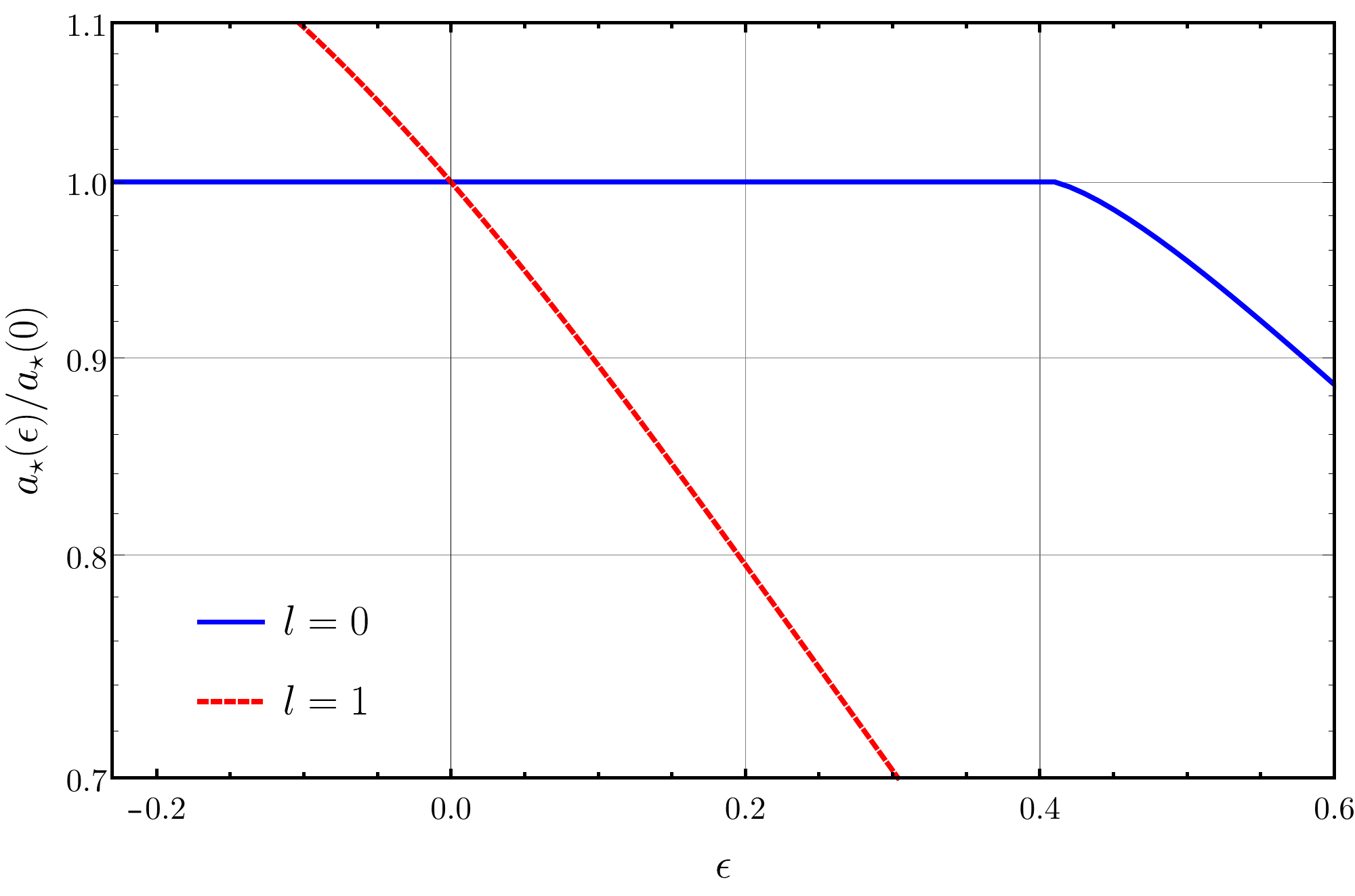}
\caption{Plots of the blow-up threshold amplitude starting from either
  a pure spherical solution (blue curve) or an~$l=1, m=0$ solution
  (red dashed curve), and adding in each case by~$l=2, m=0$ spherical
  harmonic parameterized by~$\epsilon$. See the main text following
  Eqn.~\eqref{eqn:perturbed_spherical_critical} for details. There is
  a neighborhood around the spherical solution in which the
  nonspherical deformation makes absolutely no difference to critical
  amplitude, although the asymptotic solution in the past light cone
  of the blow-up point is modified. Once the perturbation is
  sufficiently large however the blue curve does bend away. At this
  point the threshold solution takes a structure similar to that
  illustrated in Fig.~\ref{Fig:l2DSS}. We expect that when the blue
  curve is extended to the left, eventually the threshold solution
  will take the form illustrated in~\ref{Fig:Ring}. In contrast, the
  pure~$l=1$ threshold amplitude is immediately affected by the
  perturbation.}
\label{Fig:Attractor}
\end{figure}

\paragraph*{\textbf{Blow-up amplitudes under perturbations.}} The previous
example shows that threshold solutions constructed from a generic
single harmonic are not unique, and may differ even in the topology of
their blow-up. In the spherical setting we have seen that adding
arbitrary small perturbations to the initial data at the threshold
nevertheless leave us with the same critical amplitude. So an obvious
question is whether or not threshold solutions built from a single
harmonic, or sum of harmonics are locally unaffected by adding
additional harmonics in the same sense. The answer is no. To see this,
recall that the mechanism for this outcome in the spherical case was
that higher order partial wave solutions vanish at the origin, where
blow-up is guaranteed to occur with spherical symmetry. In general the
support of higher order partial waves includes however possible
blow-up points induced by another partial wave solution. Therefore a
small addition of a higher order partial wave can render a threshold
solution small enough to avoid blow-up or drive it unambiguously over
the threshold. The difference between the spherical and generic setup
is illustrated by Fig.~\ref{Fig:Attractor}. General threshold
solutions are thus described as a sum over all harmonics, with any
individual harmonic potentially playing a role in the blow-up, and
with different topologies, like the ring of Fig.~\ref{Fig:Ring}, of
the singular points possible. This behavior could be sidestepped if we
re-expanded the solution in terms of translated spherical harmonics
centered at the blow-up point, or more generally a point in the curve
of blow-up points, to again recover a basis well-adapted to the
solution at hand.

\paragraph*{\textbf{Self-similarity and generic threshold solutions.}}
By definition, a generic threshold solution can be obtained through
the deformation~$\phi_\star=D(\varphi_\star)$, where~$\varphi_\star$
is a solution of the flat-space wave equation such that
for~$t<t_\star$ we have~$\varphi_\star(t,x^i)>\xi_\star$,
and~$\varphi_\star(t_\star,x^i_\star)=\xi_\star$ is a local
minimum. Again, the point~$(t_\star,x^i_\star)$ is taken to be the
location of blow-up of the deformed solution. Because~$\varphi_\star$
is a local minimum at $(t_\star,x^i_\star)$, all first derivatives
vanish at this point, and some second derivatives must be positive,
like~$\p_t^2 \varphi_\star$; however, the second derivatives~$\p_t
\p_{i} \varphi_\star$ and~$\p_{j} \p_{i} \varphi_\star$ may be zero if
the blow-up happens in a curve or a surface (as illustrated in
Fig.~\ref{Fig:Ring}). We assume here that the blow-up happens at a
point, but the same discussion applies to any point in a curve or
surface of blow-up, with the caveat that the past light cone of each
such point can be treated locally as follows, with a global
understanding to be tackled separately. Close to this blow-up point,
the solution of the original flat-space wave equation is
\begin{align}
  \lim_{(t,x^i)\to(t_\star,x^i_\star)}\varphi_\star&\sim \xi_\star
  +\tfrac{1}{2} \p_t^2\varphi_\star(t_\star-t)^2\nonumber \\
  &-\p_t \p_{i}\varphi_\star(t_\star-t)(x^i-x^i_\star)\nonumber\\
  &+\tfrac{1}{2}\p_{i} \p_{j}\varphi_\star(x^i-x_\star^i)(x^j-x^j_\star)\,,
\end{align}
with all derivatives evaluated at~$(t_\star,x^i_\star)$. Uniqueness of
the threshold solution in the spherical case, and the lack thereof in
general, can be understood here from the fact that the derivatives in
the former case depend only on the~$l=m=0$ partial wave solution,
whereas in general higher harmonics can contribute. To count the
number of free-parameters here, first observe that, performing a
trace-trace-free decomposition on~$\p_i\p_j\varphi_\star$, the Laplace
piece can be replaced using the wave equation. We then count nine free
parameters. If we introduce a spherical harmonic decomposition
of~$\varphi_\star$ centered at~$x_\star^i$, it follows by the~$O(r^l)$
property of the partial waves that only the lowest order (up to~$l=2$)
harmonics can contribute, which gives a consistent count of
parameters. The first derivatives are
\begin{align}
  \lim_{(t,x^i)\to(t_\star,x^i_\star)}\p_ t\varphi_\star&\sim
  \p_t^2\varphi_\star(t_\star-t)
  +\p_t\p_{i}\varphi_\star(x^i-x^i_\star)\,,
  \nonumber\\
  \lim_{(t,x^i)\to(t_\star,x^i_\star)}\p_ {i}\varphi_\star
  &\sim \p_{i}^2\varphi_\star(t_\star-t)
  -\p_t \p_{i}\varphi_\star(t_\star-t)\,.
\end{align}
Let's look at the models arising from deformations using periodic
functions. Using model~$3$, for instance, which has~$\xi_\star=-1$, we
have
\begin{align}
  &\lim_{(t,x^i)\to(t_\star,x^i_\star)}\p_t\phi_{1\star}\sim \nonumber\\
  &\quad-\lim_{(t,x^i)\to(t_\star,x^i_\star)}\cos\left[A_3^{-1}\log
    \left(1+\varphi_\star\right)\right]
  \frac{\p_t \varphi_\star}{1+\varphi_\star}
\end{align}
and
\begin{align}
  &\lim_{(t,x^i)\to(t_\star,x^i_\star)}\p_{i}\phi_{1\star}\sim\nonumber \\
  &\quad-\lim_{(t,x^i)\to(t_\star,x^i_\star)}\cos\left[A_3^{-1}\log
    \left(1+\varphi_\star\right)\right]\frac{\p_{i} \varphi_\star}
  {1+\varphi_\star}\,.
\end{align}
Close to the point~$(t_\star,x^i_\star)$, the
denominator~$(1+\varphi_\star)$ is quadratic
in~$(t_\star-t,x^i-x_\star^i)$ and the first
derivatives~$\p_t\varphi_{1\star}$ and~$\p_{i}\varphi_{1\star}$ are
linear in the same argument. So, the argument applied to spherically
symmetric solutions goes through, and we conclude that the blow-up
of~$\p_t\phi_{1\star}$ and~$\p_{i}\phi_{1\star}$ is DSS, centered
at~$(t_\star,x^i_\star)$, with~$\nu=-1$ and~$\lambda_m=e^{-m
  \Delta}=e^{m\pi A_3}$. Thus we find that the CSS and DSS blow-up
properties of spherically symmetric threshold solutions, and even the
non-standard behavior with our more general compactifications like in
model~$5$, can be extended to arbitrary threshold solutions. Now,
however, nine parameters rather than one are required to characterize
the asymptotic solution in the past-light cone of the blow-up point.

\paragraph*{\textbf{Power-law scaling around general threshold solutions.}}
So far we have discussed power-law behavior that occurs in physical
space. In critical collapse such behavior is usually viewed in
phase-space. We turn our attention to this next, working with the time
derivative of the field, since this allows us to treat both types of
model in a unified way. Consider a family of
solutions~$\phi_a=D[a\,\varphi]$, parameterized by~$a$, with~$\varphi$
a fixed, nontrivial solution of the wave equation which explodes the
deformation function first at~$a=a_\star$ as usual. Let~$x^\mu(a)$ be
the locus of maxima (in amplitude) of~$\Pi_a=\p_t\phi_a$, which we
assume defines a curve when~$a\lesssim a_\star$, with~$a_\star$ the
threshold amplitude. Since~$\Pi_a$ attains a local maxima
at~$x^\mu(a)$, we have
\begin{align}
  \p_t\Pi_a=aD'(a\,\varphi)\p_t^2\varphi+a^2D''(a\,\varphi)
  (\p_t\varphi)^2=0\,,\label{eqn:TurningPoint}
\end{align}
which is understood to hold at~$x^\mu(a)$, and which we can solve
for~$(\p_t\varphi)^2$. Since this equation must hold for all values
of~$a$, we can derive in~$a$, and obtain an expression for~$t'(a)$ in
terms of the other variables. Assuming more regularity on the curve,
we are free to take higher derivatives too. Starting with the general
expression for~$\Pi_a$ we then get
\begin{align}
  \Pi_a(x^\mu(a))^{-2}&= \frac{D''(a\,\varphi)}{a D'(a\,\varphi)^3\p_t^2\varphi}
  \,,\label{eqn:oopi2ofa}
\end{align}
again understood to hold at~$x^\mu(a)$. From here we split our
discussion into two cases. First suppose that~$D=\mathcal{C}$ with our
compatification~\eqref{eqn:compactification}, assuming that~$n>0$. In
this case~\eqref{eqn:oopi2ofa} takes the form,
\begin{align}
  \Pi_a(x^\mu(a))^{-2}&=
  -\frac{\mathcal{C}''(a\,\varphi)}
       {a \mathcal{C}'(a\,\varphi)^3\p_t^2\varphi}
       =  \frac{(1+a\,\varphi)^{2n+1}}{a\,\p_t^2\varphi}
  \,.\label{eqn:oopi2ofa_compactification}
\end{align}
We need to extract the piece of this that dominates as~$a\to a_\star$.
Since~$\p_t^2\varphi$ is generically non-zero at the maximum and
non-zero as~$a\to a_\star$, we need only consider
\begin{align}
  \frac{\mathcal{C}''(a\,\varphi)}{\,\mathcal{C}'(a\,\varphi)^3}
  &=-(n+1)(1+a\,\varphi)^{2n+1}\,.
\end{align}
Raising this to the power~$1/(2n+1)$, Taylor expanding at an
arbitrary~$a=a_0$, plugging in the result for~$t'(a)$, and taking the
limit~$a_0\to a_{\star}$ we conclude that, in the regime~$a\lesssim
a_\star$, we have
\begin{align}
\Pi_a(x^\mu(a))&\simeq (a-a_\star)^{-(2n+1)/2}\,.
\end{align}
The logarithmic compactification used in~\eqref{model1_D} is more
subtle to treat, but corresponds to the case~$n=0$. In fact for this
model the range~$-1/2<n<0$ may also be interesting to investigate, but
we do not do so here. Moving now to the
case~$D=\mathcal{P}\circ\mathcal{C}$, again for concreteness taking
the compactification from~\eqref{eqn:compactification}, we find
that~\eqref{eqn:oopi2ofa_compactification} is instead replaced by
\begin{align}
  -\Pi_a(x^\mu(a))^{-2}&=\frac{\mathcal{P}''}
     {a\mathcal{P}'^3\mathcal{C}(a\,\varphi)\p^2_t\varphi}
     +\frac{\mathcal{P}'\mathcal{C}''(a\,\varphi)}
     {a\mathcal{P}'^3\mathcal{C}(a\,\varphi)^3\p^2_t\varphi}\,.
\end{align}
Following from here the same procedure as before, noting that the
first of these terms is now the leading piece, and raising to the
power~$1/(n+1)$, in the regime~$a\lesssim a_\star$, we find that
\begin{align}
\Pi_a(x^\mu(a))&\simeq (a-a_\star)^{-(n+1)/2}\,.
\end{align}
Again the~$\log$ compactification can be thought of as~$n=0$. With a
little more care we expect that one could see here also the superposed
periodic wiggle. An important message here is that power-law behavior
may appear even in models for which self-similarity is absent at the
threshold, so evidence of both phenomena are needed for a confident
diagnosis. In summary, we find that under mild assumptions on the
regularity of~$x^\mu(a)$, close to the threshold, all of our models
admit universal power-laws regardless of the nature of the threshold
solution itself. Nevertheless some care is needed in interpreting this
result. For general data there may appear multiple ``large-data''
regions, and the peak of that which ultimately leads to blow-up in the
limit~$a\to a_\star$ may be obfuscated, over some range of~$a$, by
another.

\paragraph*{\textbf{Regularity of threshold vs. generic blow-up solutions.}}
Recovering results on the norms of threshold and blow-up solutions in
the nonspherical setting is trickier than the previous case. Although
the only numerical part of the calculation is in the evaluation of the
norm itself, the solutions can be highly oscillatory. Nevertheless in
all of the cases that we can reliably verify, which include all of
those presented in Fig.~\ref{Fig:NormPlot}, we find that our spherical
results carry over without any surprises, and that threshold solutions
are slightly more regular than generic blow-up solutions. In the
future it will be interesting to use the geometric reformulation of
our models given in Sec.~\ref{Section:Models} together with the
conserved stress-energy to prove these properties beyond doubt.

\section{Numerical Results}\label{Section:Numerics}

In the previous section we gave a fairly complete picture of threshold
solutions for the models that arise as a deformation of the wave
equation. To address the obvious criticism that such models may not be
qualitatively representative of systems that do not arise as a
deformation, we now present numerical evidence that similar
phenomenology does occur within our non-deformation models. Presently
we restrict to spherical symmetry, postponing detailed numerical of
generic threshold solutions for future work. We begin by explaining
briefly the method used, before presenting the classification and
numerical results for each model. Similar, though more comprehensive,
numerical work for alternative models can be found
in~\cite{Lie02,BizChmTab04,Lie05a,BizMaiWas07,BizZen08}.

\subsection{Methods}\label{Subsection:Method}

As presented in section~\ref{Section:Models}, all model equations are
second order both in time and space. For the code we reduce the system
to fully first order form and use centered finite differences. To do
so we introduce the following auxiliary evolved fields,
\begin{align}
\Phi = \p_r \phi\,,\qquad\qquad \Pi = \p_t \phi\,.
\label{eqn:Auxiliar_fields}
\end{align}
In order to deal with the coordinate singularity at the origin, we
apply the Evans method, for any scalar field~$\Psi$ and its
derivative~$\Psi' = \frac{d \Psi}{dr}$, with~$p = 2$~\cite{Eva84},
\begin{align}
\Psi' + \tfrac{p}{r} \Psi = (p + 1) \frac{d(r^p \Psi)}{d(r^{p+1})}\,,
\label{eqn:EvansMethod}
\end{align}
where the differential operator can be expressed in terms of the grid
points as,
\begin{align}
  (p + 1) \frac{d(r^p \Psi)}{d(r^{p+1})} = (\tilde{D}\Psi)_i = (p + 1)
  \frac{r^p_{i+1} \Psi_{i+1} - r^p_{i-1} \Psi_{i-1}}{r^{p+1}_{i+1} -
  r^{p+1}_{i-1}}.
\label{eqn:EvansOperator}
\end{align}
In section~\ref{Section:Self-similar} definitions for the different
norms we consider were given, and their blow-up for CSS and DSS
functions was introduced and related. Below in this section we
classify the models presented in section~\ref{Section:Models}
following this criteria. We employ the method of lines with a
Runge-Kutta~$4$ time integrator, and to avoid rapid growth of
numerical error we use second order Kreiss-Oliger artificial
dissipation~\cite{KreOli73} with a small dissipation parameter of
order~$\sigma = 0.02$. The particular boundary conditions for each
model are stated at their corresponding section.

\subsection{CSS and $L^{\infty}$ blow-up}\label{Subsection:CSS}

\begin{figure}[t!]
\centering
\includegraphics[width=0.5\textwidth]{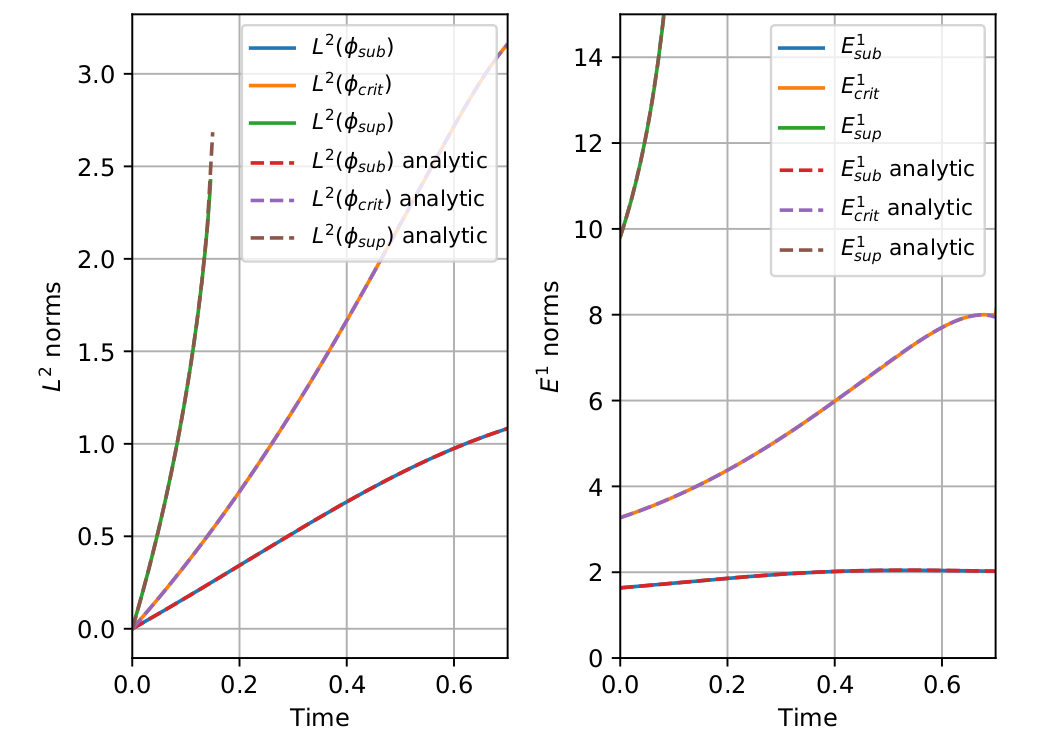}
\caption{$L^2$ and~$E^1$ for model~$1$ for sub, critical and
  supercricital data computed from our numerical simulations and the
  exact solution for model~$1$ with~$A_1=-1$. The numerical data agree
  extremely well with the values computed from the exact
  solution. This indicates that, with suitable care, numerical
  evolutions can be of real value in determining regularity even at
  blow-up.}
\label{Fig:M1_Norms}
\end{figure}

\paragraph*{\textbf{Model~$\boldsymbol{1}$.}} The equations of motion for
the auxiliary fields are,
\begin{align}
\p_t \Phi &= \p_r \Pi\,, \nonumber\\
\p_t \Pi &= \p_r \Phi
+ \tfrac{2}{r} \Phi + A_1 \left(\Phi^2 - \Pi^2 \right).
\label{eqn:Motion_M1}
\end{align}
We impose the condition
\begin{align}
\p_t \Pi  \; \hat{=} - \p_r \Pi,
\label{eqn:Boundary_M1}
\end{align}
at the outer boundary. Modulo boundary effects which are negligible in
our present study, we can write down closed form solutions for this
model, so numerical work constitutes only a code-test. But such work
can be highly valuable as it may give confidence in purely numerical
studies and highlight algorithmic shortcomings. We have performed
numerical evolutions with a variety of initial data and find that the
method converges reliably at second order as expected. As observed in
section~\ref{Section:Critical} this model is an example with
approximately CSS threshold behavior. All blow-up solutions, including
those at the threshold, explode in~$L^\infty$, but nevertheless {\it
  may} remain finite and even in~$L^2$ and even in the energy
norm~$E^1$. At the threshold solutions are finite in~$E^1$, whereas
generically blow-up solutions explode in~$E^1$. An important question
therefore is how well numerical methods can cope with data at these
varying levels of regularity. Pessimistically one might expect that
with standard methods when the solution explodes pointwise, numerical
error becomes large so fast that any approximation to~$L^2$ (and so
forth) from the numerical data also diverges. We have investigated
this, as shown for example in Fig.~\ref{Fig:M1_Norms}, and find that
the numerics capture the expected behavior well. In the future it may
be useful to examine the same question for models that have different
regularity at blow-up, for example by using our parameterized
compactification~\eqref{eqn:compactification}.

\begin{figure*}[t] 
\centering
\includegraphics[width=0.47\textwidth]{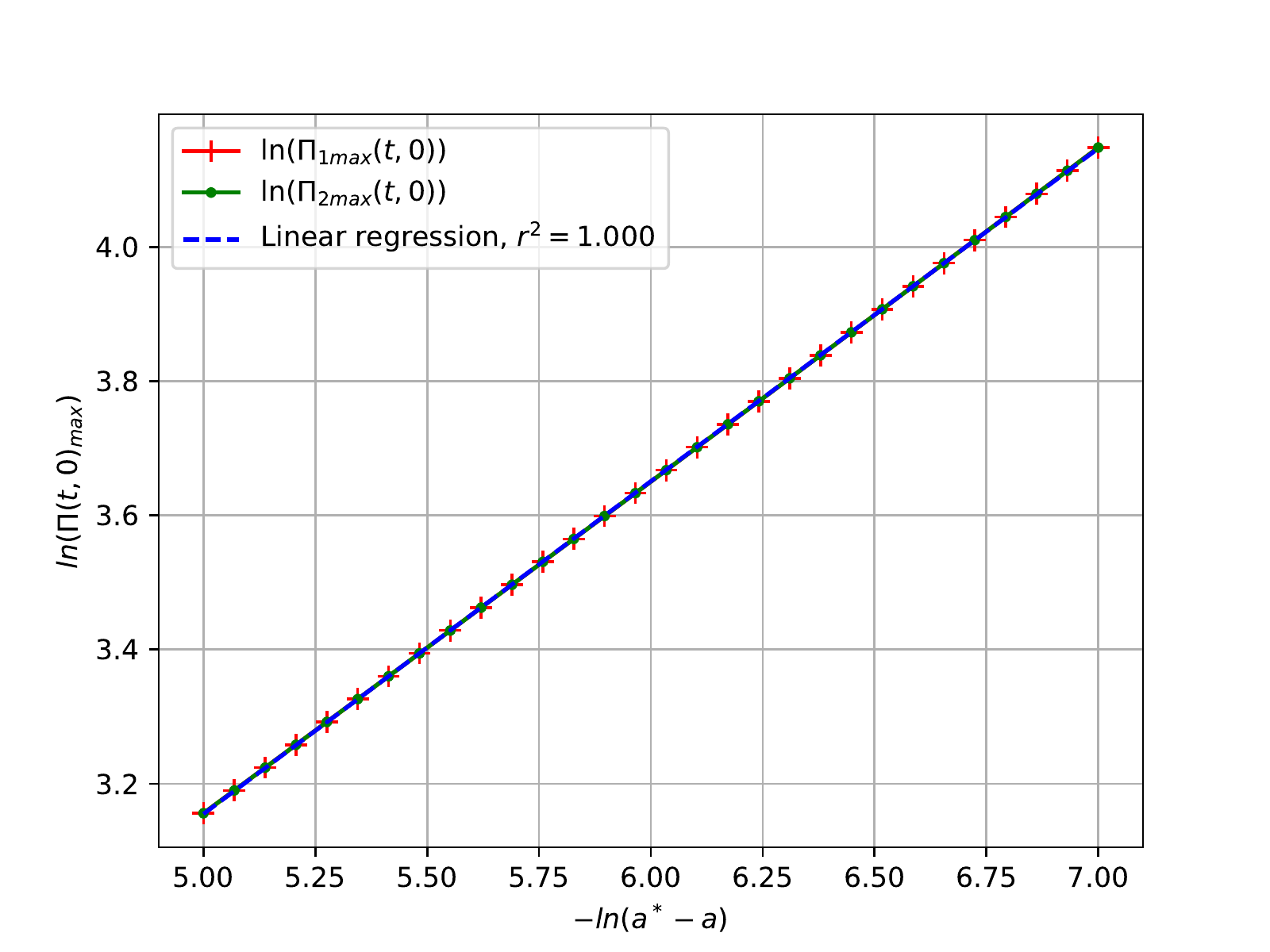}
\includegraphics[width=0.47\textwidth]{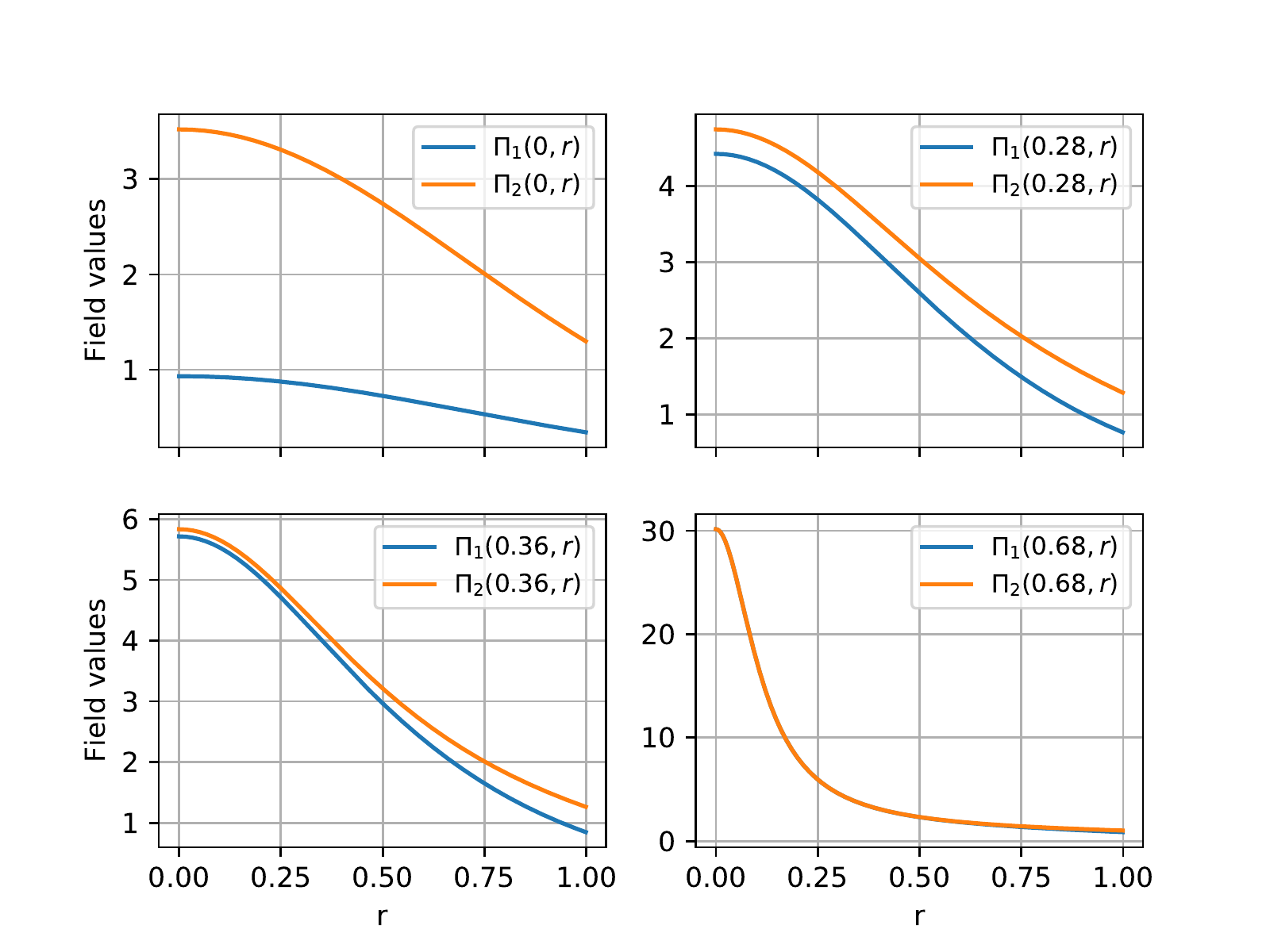}
\caption{In the left panel we plot the scaling law obtained close to
  the threshold by taking the maximum of the time derivatives of the
  evolved fields~$\phi_1, \phi_2$ for Model~$2$. We have
  chosen~$A_2=B_2=-1$, and used initial data as stated
  in~\eqref{eqn:ID_M2}. The threshold amplitude~$a_\star = 1.5103468$
  was obtained by numerical bisection. In the legend~$r^2$ refers to
  the square of the Pearson correlation coefficient, which we computed
  using the Scipy Python library~\cite{VirGomOli20}. A best fit on the
  data at this resolution returns the gradient~$0.49594$ with standard
  error~$0.00018$. On the right we plot snapshots of the same fields
  close to blow-up for the threshold solution itself. Observe that the
  fields lie on top of each other in the at late times, indicating
  that the threshold solution is in fact described by the same
  critical solution of model~$1$. Identical results are obtained with
  other families of initial data.\label{Fig:M2} }
\end{figure*}

\paragraph*{\textbf{Model~$\boldsymbol{2}$.}} The equations of motion for
the reduction variables are
\begin{align}
  \p_t \Phi_1 &= \p_r \Pi_1\,, \qquad \p_t
  \Phi_2 = \p_r \Pi_2 \nonumber\,, \\
  \p_t \Pi_1 &= \p_r \Phi_1 + \tfrac{2}{r} \Phi_1 + A_2 \left( \Pi_2^2 -
  \Phi_2^2 \right) \,, \nonumber\\
  \p_t \Pi_2 &= \p_r \Phi_2 + \tfrac{2}{r} \Phi_2
  + B_2 \left( \Pi_1^2 - \Phi_1^2\right)\,.\label{eqn:Motion_M2}
\end{align}
At the outer boundary we impose
\begin{align}
  \p_t \Pi_{1} \; \hat{=} - \p_r \Pi_{1}\,, \qquad
  \p_t \Pi_{2} \; \hat{=} - \p_r \Pi_{2}  \,.
\label{eqn:Boundary_M2}
\end{align}
We have evolved and tuned to the threshold of blow-up with several
families of initial data, but here discuss a representative example
with initial data,
\begin{align}
  \Phi_1(0,r)&=\Phi_2(0,r) = 0\,,\nonumber\\
  \Pi_1(0,r) &= \tfrac{2}{5}e^{1/2 - r^2}\,,\quad
  \Pi_2(0,r) = a e^{1/2 - r^2}\,.\label{eqn:ID_M2}
\end{align}
We have experimented with various choices for the parameters~$A_2$
and~$B_2$, which do not seem to affect the qualitative behavior of
solutions. Recall that if we choose~$A_2 =B_2=A_1$ and set~$\phi_1 =
\phi_2$ we recover solutions of model~$1$, making this choice of the
parameters an interesting point to investigate in more detail. In
Fig.~\ref{Fig:M2} we do so by plotting the logarithm of the maximum of
the time derivative of the scalar field at the
origin~($\Pi_1(t,0)_{\textrm{max}}$, $\Pi_2(t,0)_{\textrm{max}}$)
against the logarithmic distance to the critical point~$a_\star$
together with their respective linear least-squares regressions. Note
that hereafter~$a$ is the only parameter in each family of solutions
and~$a_\star$ refers to its critical value in each case. Note that
there are two lines, one red and one green, but near the threshold
they perfectly overlap and give, as a result, the figures mentioned
above. Interestingly, in fact we find that for any strong data,
with~$A_2=B_2$, the two sets~$(\phi_1,\Phi_1,\Pi_1)$
and~$(\phi_2,\Phi_2,\Pi_2)$ miraculously coincide, and so in fact
threshold solutions agree with those of model~$1$. This behavior is
shown in the right panel of Fig.~\ref{Fig:M2}. Scaling shows that
if~$A_2B_2>0$ then,
\begin{align}
A_1^{-1}(A_2B_2^2)^{1/3}\phi_1\,,\qquad A_1^{-1}(A_2^2B_2)^{1/3}\phi_2\,,
\end{align}
solve the same model with fresh constants~$A_2'=B_2'=A_1$. This is of
course borne out in our simulations. Our numerical evidence therefore
strongly suggests that all spherical threshold solutions can be
constructed directly from model~$1$. We conclude that model~$2$ does
indeed have a unique critical solution in spherical symmetry. Given
this, it is perhaps not surprising that experiments indicate the same
level of regularity in~$L^2$ and~$E^1$ for this model as for model~$1$
in Fig.~\ref{Fig:M1_Norms} for subcritical, critical and supercritical
initial data.

\subsection{DSS models and their blow-up}\label{Subsection:DSS}

\begin{figure}[t] 
\centering
\includegraphics[width=0.47\textwidth]{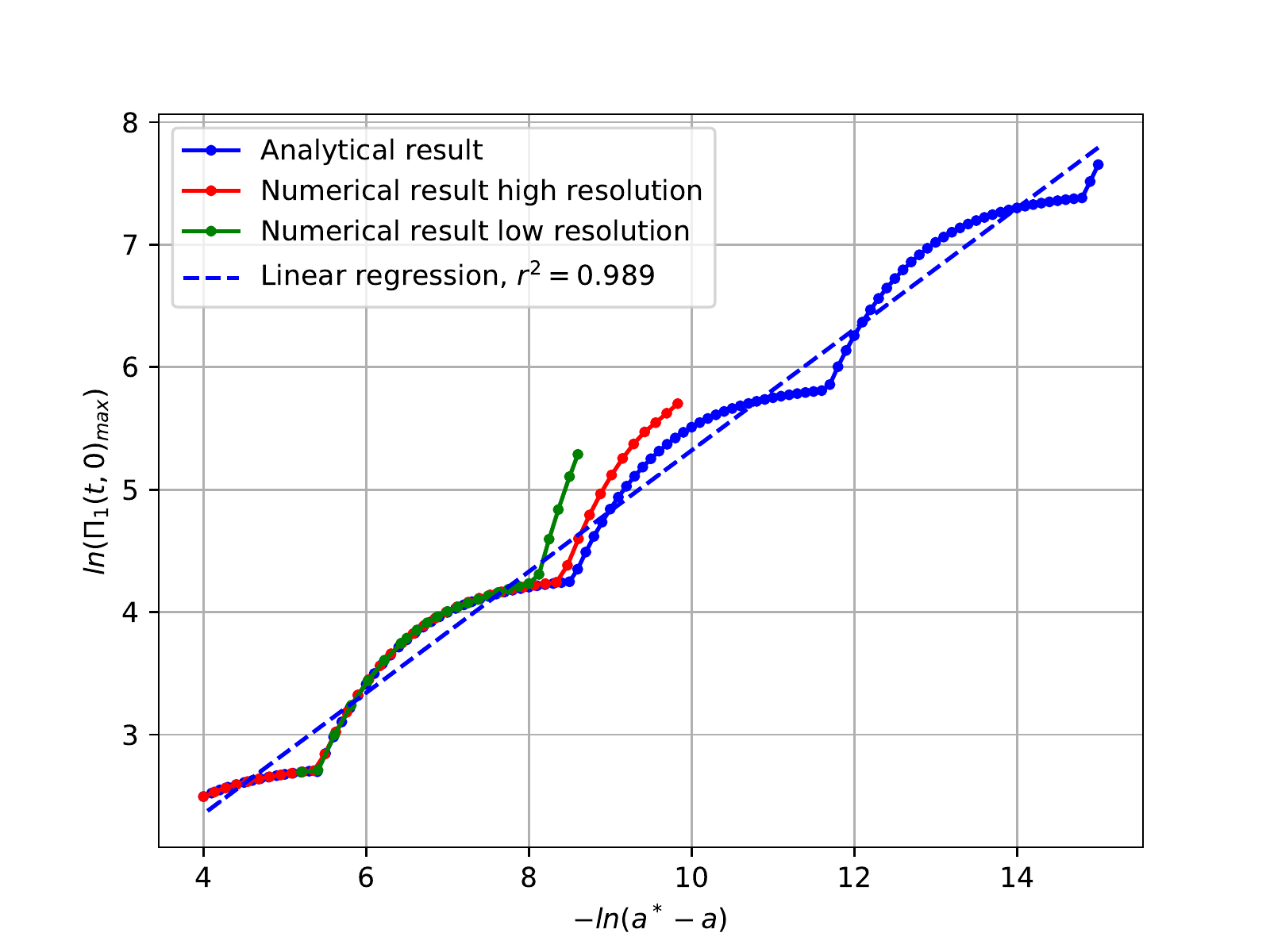}
\caption{Scaling plot for~$\Pi_1(t,0)_{\textrm{max}}$ for model~$3$
  with~$A_3=1$ for the family of initial data~\eqref{eqn:ID_M3}. The
  threshold amplitude for this family is~$a_\star=-\sqrt{2}$. For
  comparison the analytic result is also given. The drift between the
  numerical and analytic curves is caused by numerical error, but does
  converge away with resolution, as can be understood from the higher
  resolution data. In the legend~$r^2$ again refers to the square of
  Pearson correlation coefficient, which was computed from the lower
  resolution data and is close to unity. Linear regression on the
  numerical data gives the gradient~$0.4945$, with standard
  error~$0.0049$, close to the expected value~$1/2$ seen in
  Sec.~\ref{Section:Critical}. \label{Fig:Scaling_M3}}
\end{figure}

\paragraph*{\textbf{Model~$\boldsymbol{3}$.}} The equations of motion
for the third model,
\begin{align}
  \p_t \phi_1 &= \Pi_1\,, \quad\quad
  \p_t \phi_2 = \Pi_2\,,\nonumber\\ 
  \p_t \Phi_1 &= \p_r \Pi_1\,,\quad 
  \p_t \Phi_2 = \p_r \Pi_2\,, \nonumber\\
  \p_t \Pi_1 &= \p_r \Phi_1 + \tfrac{2}{r} \Phi_1\nonumber\\
  &\quad+ A_3^{-2}( \phi_1 + A_3 \phi_2 )
  [ \Phi_1^2 + \Phi_2^2 - \Pi_1^2 - \Pi_2^2 ]
  \,, \nonumber\\
  \p_t \Pi_2 &= \p_r \Phi_2 + \tfrac{2}{r} \Phi_2\nonumber\\
  &\quad+ A_3^{-2}( \phi_2 - A_3 \phi_1 )
  [ \Phi_1^2 + \Phi_2^2 - \Pi_1^2 - \Pi_2^2 ] \,,
  \label{eqn:Motion_M3}
\end{align}
are supplemented with the corresponding boundary conditions,
\begin{align}
  \p_t \Pi_1 \; \hat{=} - \p_r \Pi_1 -
  \tfrac{1}{r}\Pi_1\,, \quad
  \p_t \Pi_2 \; \hat{=} - \p_r \Pi_2
  - \tfrac{1}{r}\Pi_2\,.
\label{eqn:Boundary_M3}
\end{align}
These boundary conditions are modified with respect to those of the
previous models simply to avoid code crashes, but in all applications
we nevertheless keep the outer boundary causally disconnected from the
region at the center we are actually interested in. Like model~$1$, we
know form solutions here, and so view these numerics primarily as a
code-test. In this spirit, in Fig.~\ref{Fig:Scaling_M3} we again show
the logarithm of the maximum of the time
derivative~$\Pi_1(t,0)_{\textrm{max}}$ against the logarithmic
distance to the critical point for a representative family of initial
data given by
\begin{align}
\phi_1(0,r) &= 0\,,&\quad \phi_2(0,r) = 1\,,\nonumber\\
\Phi_1(0,r) &= 0\,,&\quad \Phi_2(0,r) = 0\,,\nonumber\\
\Pi_1(0,r) &= a e^{1/2 - r^2}\,,&\quad \Pi_2(0,r) = 0\,,\label{eqn:ID_M3}
\end{align}
in this instance using~$A_3=1$. In all cases we clearly observe the
expected DSS behavior, which manifests as a straight line plus a
periodic wiggle whose period depends on the value of~$A_3$. Regarding
regularity, recall that this model actually has the similar behavior
as model~$1$. Although the solution itself never diverges, first
derivatives are divergent for any blow-up solution. Solutions are
always finite in~$L^2$. At the threshold~$E^1$ is finite, but for all
other blow-up solutions it diverges. We have examined how well this
behavior is captured in our numerical approximation and find that
results similar to those displayed in Fig.~\ref{Fig:M1_Norms} are
easily obtained, albeit with~$L^2$ finite, and that these results
agree very well with those computed directly from the exact solution,
even at blow-up.

\begin{figure*}[t] 
\centering
\includegraphics[width=0.47\textwidth]{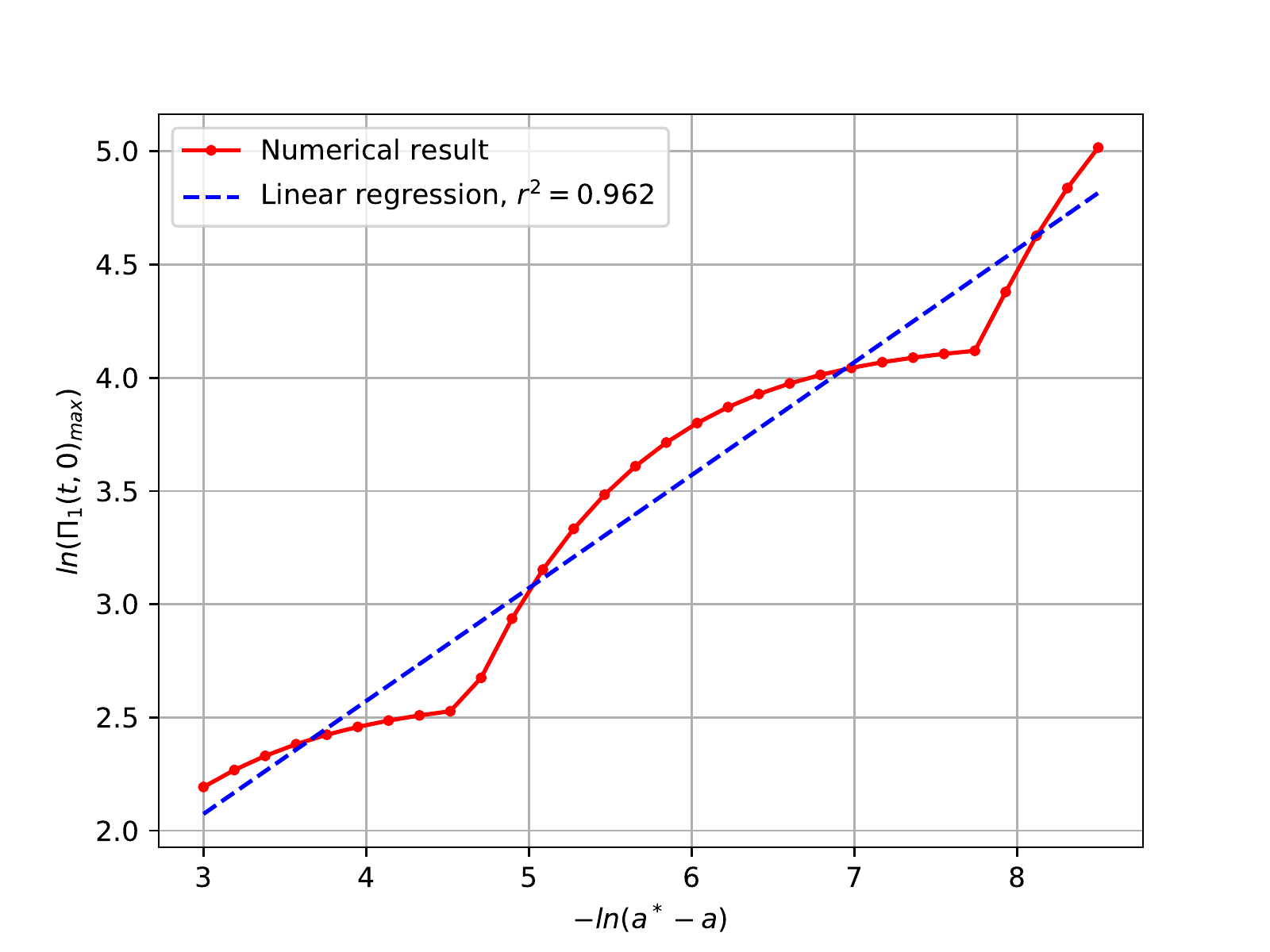}
\includegraphics[width=0.47\textwidth]{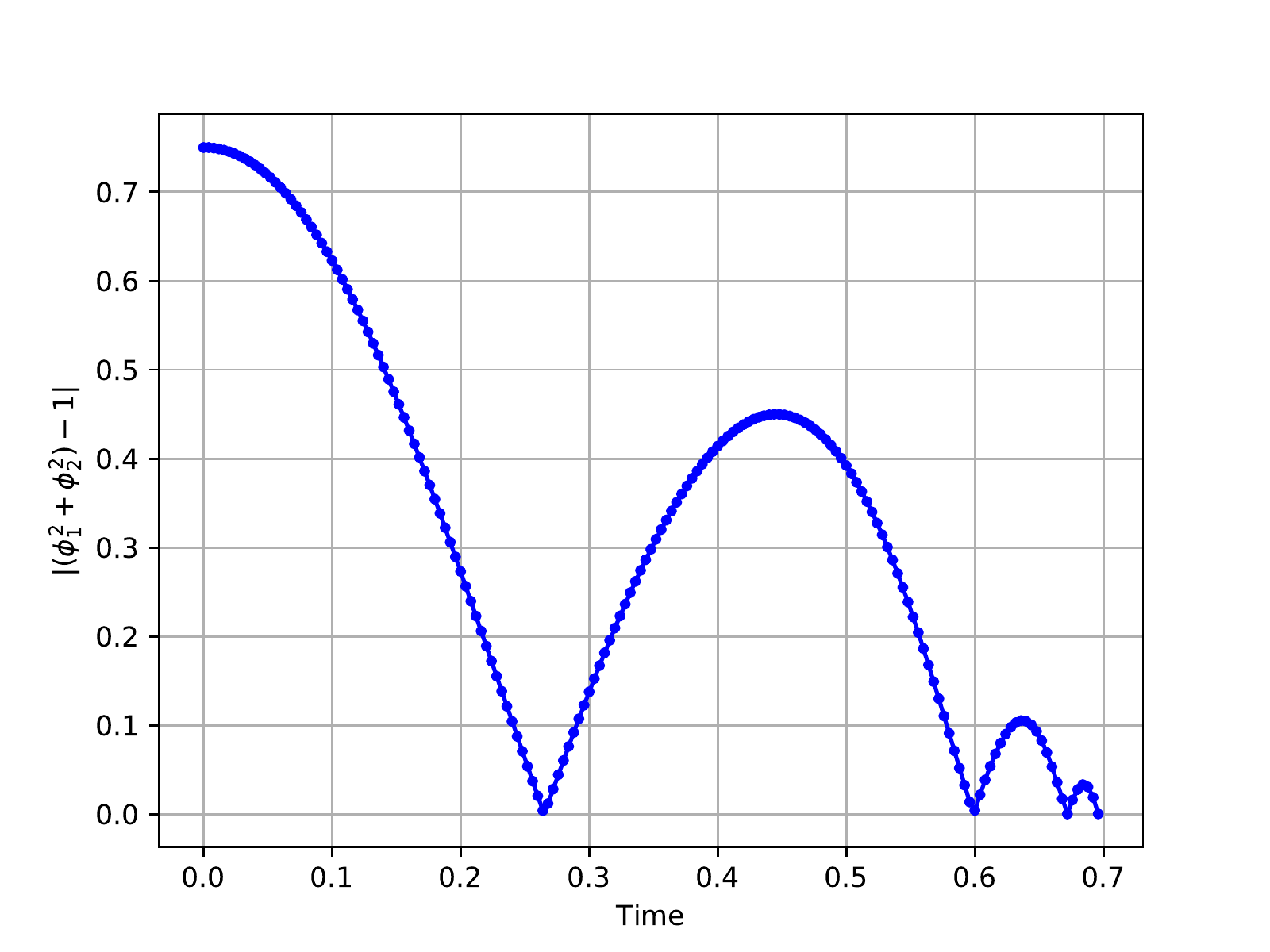}
\caption{Representative plots obtained with model~$4$ with~$A_4=B_4=1$
  and the initial data family~\eqref{eqn:ID_M4}. This is obtained
  with~$a_\star\simeq-2.4122$. On the left we give the now familiar
  scaling plot for~$\Pi_1$. As in model~$3$ the curve looks like a
  straight line plus a periodic wiggle, indicating that we are in a
  DSS regime. Linear regression on the numerical data gives a
  slope~$0.499$, with standard error~$0.019$. On the right we plot the
  maximum of the absolute value at the origin of the quantity that
  serves as a constraint in model~$3$. In fact this quantity is small
  in a neighborhood around the origin, so that near the threshold,
  solution of model~$4$ are close to solutions of
  model~$3$.\label{Fig:M4}}
\end{figure*}

\paragraph*{\textbf{Model~$\boldsymbol{4}$.}} The final model that we
implemented is an extension of model~$3$ in which the
constraint~$\phi_1^2+\phi_2^2=A_3$ is violated. The equations of
motion in the case are,
\begin{align}
  \p_t \phi_1 &= \Pi_1\,, \quad\quad
  \p_t \phi_2 = \Pi_2\,,\nonumber\\ 
  \p_t \Phi_1 &= \p_r \Pi_1\,,\quad 
  \p_t \Phi_2 = \p_r \Pi_2\,, \nonumber\\
  \p_t \Pi_1 &= \p_r \Phi_1 + \tfrac{2}{r} \Phi_1\nonumber\\
  &\quad+ A_4^{-2}( \phi_1 + A_4 \phi_2 )
  [ \Phi_1^2 + \Phi_2^2 - \Pi_1^2 - \Pi_2^2 ]
  \,, \nonumber\\
  \p_t \Pi_2 &= \p_r \Phi_2 + \tfrac{2}{r} \Phi_2\nonumber\\
  &\quad+ B_4^{-2}( \phi_2 - B_4 \phi_1 )
  [ \Phi_1^2 + \Phi_2^2 - \Pi_1^2 - \Pi_2^2 ] \,.
  \label{eqn:Motion_M4}
\end{align}
In this case, the two scalar fields of the model are not, a priori,
related to each other because solutions do not arise from a
deformation of the wave equation. In Fig.~\ref{Fig:M4} we plot the
logarithm of the maximum of the time
derivative~$\Pi_1(t,0)_{\textrm{max}}$ against the logarithmic
distance to the critical point and observe that this model, despite
violating the constraint and not coming from a deformation of the wave
equation, exhibits DSS behavior too. In this particular plot we worked
with~$A_4 = B_4 = 1$, and the family of initial data,
\begin{align}
\phi_1(0,r) &= 0\,, &\quad \phi_2(0,r) = \tfrac{1}{2}\,,\nonumber\\
\Phi_1(0,r) &= 0\,, &\quad\Phi_2(0,r) = 0\,,\nonumber\\
\Pi_1(0,r) &= a e^{1/2 - r^2}, &\quad\Pi_2(0,r) = 0\,.\label{eqn:ID_M4}
\end{align}
and tuned to the threshold~$a_\star = -2.4122175$ by numerical
bisection. Similar to model~$2$, close to the threshold we observe
that, at least for the families of data that we tested, the
``constraint'' is in fact small close to criticality. We observe
similar behavior for any blow-up solution, but it is most pronounced
at the threshold. This is illustrated in the second plot of
Fig.~\ref{Fig:M4}. We do note however, that this behavior is not as
striking as in model~$2$, where the ``constraint'' seems identically
satisfied over an entire region, rather than just being small as in
this case. Concerning regularity, at the threshold the raw
fields~$\phi_1$ and~$\phi_2$ remain finite (and thus the solution
remains finite in~$L^2$), but as shown in the discussion above first
derivatives do explode. Our data suggest that the energy norm~$E^1$ is
finite at the threshold but diverges for supercritical solutions, in
agreement with model~$3$. Having examined several different families
of initial data, our numerical evidence again suggests that in
spherical symmetry model~$4$ has a unique critical solution in the
same sense as our other models.

\section{Conclusions}\label{Section:Conclusion}

The cosmic censorship conjectures are perhaps the most important open
problems in strong-field gravity. In looking for evidence either for
or against them it is imperative that we examine extreme regions of
the solution space. Combining such considerations with numerical
approximation, critical phenomena in gravitational collapse have been
discovered. The standard picture of critical collapse is that, if we
consider any one-parameter family of initial data and tune that
parameter to the threshold of black hole formation, then as it heads
towards blow-up the resulting threshold solution will approximate ever
more closely, in the strong-field region, a unique self-similar {\it
  critical} solution which has a naked singularity. In suitable
coordinates data within the family, but close to the threshold,
approach the critical solution for some time
interval~$\sim-\gamma^{-1}\log|a-a_\star|$ before either dispersing or
collapsing, with~$\gamma$ a universal parameter independent of the
particular family. Examining solutions parametrically in a
neighborhood of the threshold reveals that the curvature scalars,
black hole masses and so forth display power-law behavior, with
power~$\gamma$, in~$a_\star-a$.

In spherical symmetry numerical evidence in favor of this picture is
pristine, and there is even a proof~\cite{ReiTru19} that the Choptuik
critical solution, with the posited discrete self-similarity,
exists. Part of this phenomenology remains robustly without symmetry,
but cracks have appeared in the picture. Prominent examples are given
by the variability of the scaling parameters and apparent
contradiction of uniqueness of the critical solution in scalar field
collapse when large aspherical perturbations are
present~\cite{ChoHirLie03,Bau18}, the seeming absence of a unique
self-similar critical solution in the collapse of the electromagnetic
fields~\cite{BauGunHil19} and the consistent challenge in treating
threshold solutions in vacuum
gravity~\cite{HilBauWey13,HilWeyBru15,HilWeyBru17} and so in
recovering the results of~\cite{AbrEva93}. In all of these cases
however, we are reaching to the edge of what is possible with present
numerical methods, so there are arguments against adjusting the
standard picture until numerical error could be reliably controlled.

In the present study we therefore sought a way to side-step these
difficulties by constructing the absolute simplest {\it school-boy}
model that could capture the qualitative behavior of interest. Our
models are based on a trick of Nirenberg, admit a small-data global
existence result, and in most cases be solved analytically, making
interpretation of threshold solutions unambiguous, regardless of
symmetry. We call these deformation models. In contrast with earlier
models, they also have the advantage, at least from the point of view
of gravitation, that their nonlinearity appears in first derivatives
of the fields, just as in GR nonlinearities are of the
form~``$\Gamma^2-\Gamma^2$''. To the best of our knowledge we have
also given the first such model that admits discretely self-similar
solutions. (Other examples with such solutions are known~\cite{Tao16}
but require a large number of fields). Although the models can be
reformulated in a natural way that introduces a non-trivial spacetime
metric, they are nevertheless fundamentally tied to the flat-metric,
and so should not be thought of as a model for weak cosmic censorship.
Rather, at best we can hope to capture the properties required for
strong cosmic censorship in terms of regularity at blow-up and of
course those of critical collapse. Our findings, conclusions and
conjectures can be split into categories discussed in turn in the next
paragraphs.

\paragraph*{\textbf{Spherical symmetry.}} Restricting to {\it pure}
spherical symmetry, the obvious analog of the standard picture of
critical collapse was completely vindicated for all of our models
regardless of how they arose. For our deformation models, simple
Taylor expansion shows that generically {\it at most} one number from
the initial data survives to parameterize the threshold solution near
the blow-up point. In fact there is a measure-$0$ special case in
which this parameter vanishes, but we have not investigated this in
detail. We {\it define} this one-parameter family of Taylor expanded
threshold solutions to be the critical solution. In that one parameter
remains, it is unique in the same sense the Schwarzschild is the
unique static vacuum solution. Extracting this parameter in any
numerical setup seems impractical, however. For models that do not
arise as a deformation of the wave equation, we tackled the spherical
setting numerically and found evidence compatible with this
picture. With either type of model we found that universal power-law
behavior, for example in the maximum of any divergent field quantity,
like for example energy density, was manifest. This was shown
analytically for the deformation models. Moving on to consider small
aspherical perturbations, to avoid having to perform more costly
numerics we studied only deformation models. We found that the
critical amplitude remains fixed, and that the blow-up itself is still
dominated by the lowest spherical harmonic. From a purely mechanical
point of view, this is a simple consequence of the fact that
aspherical partial wave solutions all vanish at the
origin. Nevertheless the asymptotic threshold solution, which
maintains the scale-symmetry from the spherical setting, is deformed
as perturbations are added, perhaps in contradiction expectations, so
that a larger number of parameters are needed for its
description. Power-law scaling in this regime, both in the physical
and phase space pictures, also remains universal. The agreement with
the standard picture of critical collapse in the regime in which
numerical results are unambiguous, is striking. This gives us
confidence that our models do capture qualitatively the phenomena of
interest, and potentially do have predictive power for GR.

\paragraph*{\textbf{Strong cosmic censorship.}} As mentioned in the
introduction, the strong cosmic censorship conjecture may be thought
of as the requirement that for generic initial data the resulting
solution, when maximally extended, is unique. In the context of
blow-up, typically in the context of black hole interiors as
in~\cite{PoiIsr90,CarCosDes17}, this is taken to mean that at a Cauchy
horizon, or more generally in the limit towards any the end-point of
any incomplete geodesic, the metric should lose enough regularity that
the solution can not be extended beyond the blow-up, even if we allow
weak solutions. If this fails to be the case, perhaps by choosing
fresh data at the singular surface, we may obtain many inequivalent
extensions and so violate global uniqueness. The specific requirement
in GR~\cite{Chr99} is that there exist no coordinates in which the
Christoffel symbols are locally~$L^2$. The natural analog for our
models is the requirement that, at blow-up, solutions explode in the
energy norm~$E^1$. The conclusion from our models is that for each
type of model there exists a direct, specific, relationship between
the physical and phase space power-law parameters~$\nu$ and~$\gamma$,
and the regularity of data at blow-up. We find that threshold
solutions are more regular than generic blow-up solutions, and so
depending on the values of these parameters solutions could be
extended beyond the blow-up point. We have not investigated this in
detail, and this result may have no {\it direct} counterpart in GR,
but if it does it will permit numerical simulations a new say on
strong cosmic censorship in a variety of scenarios.

\paragraph*{\textbf{The threshold of blow-up.}} When considering either
large aspherical deformations of spherical threshold solutions or
general threshold solutions we depart from the standard picture of
critical collapse. But from an empirical point of view, our results in
this regime are nevertheless compatible with numerical results in
GR. First, power-law scaling persists both in physical space near the
blow-up point, and also in phase space as the threshold is
approached. In GR there is evidence, in scalar field collapse, that
power-law rates deviate from their values in spherical symmetry as
large asphericity appears~\cite{ChoHirLie03,Bau18} so this is a
possible difference to the models. That said, it is not obvious that
the available numerical data are sufficiently fine-tuned to recover
the limiting rates, and the interactions of multiple fields complicate
the interpretation. If spherical data for the models are perturbed by
a sufficiently large asphericity, blow-up occurs away from the origin,
with the solution displaying appearing very different to the spherical
critical solution in the past light cone of the blow-up point, in
contradiction with the expectation that there exist a unique {\it
  critical solution} in the general setting. This may manifest, for
example by the formation of multiple nonspherical centers away from
the origin. The latter has been observed in GR in both scalar
field~\cite{ChoHirLie03,Bau18} and vacuum
collapse~\cite{HilWeyBru17}. As illustrated in Fig.~\ref{Fig:Ring}
blow-up can even occur on curves rather than points, an important
possibility to be investigated in the gravitational context. Depending
on the model, general threshold solutions may exhibit self-similarity,
but require several parameters to describe them as they approach
blow-up. In GR, by analogy, the existence of a single critical
solution would be a red-herring in general. Instead, the threshold of
collapse should be characterized by power-law scaling, and, crucially,
additional regularity with respect to general blow-up
solutions. Recalling that we have models which display these features,
but do not satisfy the formal definition of self-similarity at the
threshold, and the lack of {\it exact} self-similarity in nonspherical
numerical work for GR, we conjecture that in the past light cone of a
blow-up point, threshold solutions in GR can still be described by a
finite number of parameters. In this way, we can still use the
language {\it critical solution}, but that solution must now be
thought of as a parameterized family, whose specific nature is, for
now, uncertain.

\paragraph*{\textbf{Future work.}} The study of model problems can never
definitively solve problems in the full generality that we would wish.
Several possibilities present themselves for future
developments. Regarding our models, it is highly desirable to develop
tools to rigorously prove, without using the exact solutions, the
properties of the solutions we have uncovered, and to satisfactorily
explain what are the structural conditions that determine either CSS
or DSS behavior at the threshold.  Obvious directions for numerical
work are to compute threshold solutions to the models without
symmetry, and to examine carefully whether or not the solution space
of GR exhibits the properties suggested, but as yet verified, by our
models. As mentioned in the introduction, a key shortcoming of all the
models we worked with here is that they are fundamentally semi-linear,
and thus admit no notion of black hole formation. Therefore the
construction of more sophisticated models without this shortcoming
must also be a priority.  Progress on these fronts will be reported
elsewhere.

\acknowledgments

We are grateful to Thomas Baumgarte, Piotr Bizo\'n, Edgar Gasperin,
Carsten Gundlach and Steve Liebling for helpful discussions and
guidance. DH also gratefully acknowledges support offered by IUCAA,
Pune, where part of this work was completed. The work was partially
supported by the IDPASC program PD/BD/135434/2017, the FCT (Portugal)
IF Program~IF/00577/2015, PTDC/MAT-APL/30043/2017,
Project~No.~UIDB/00099/2020, FCT PhD scholarship SFRH/BD/128834/2017
and the GWverse COST action Grant No. CA16104.

\bibliography{BondiHyp.bbl}{}

\end{document}